\documentclass[preprint,3p,10pt]{elsarticle}
\usepackage{hyperref}

\usepackage[T1]{fontenc}
\usepackage[utf8]{luainputenc}
\usepackage{verbatim}
\usepackage{amsmath}
\usepackage{amssymb}
\usepackage{graphicx}
\usepackage{esint}
\usepackage{times}
\usepackage{color}
\journal{CR Physique}

\begin{document}
\let\vaccent=\v{
\global\long\def\gv#1{\ensuremath{\mbox{\boldmath\ensuremath{#1}}}}
\global\long\def\uv#1{\ensuremath{\mathbf{\hat{#1}}}}
\global\long\def\abs#1{\left| #1 \right|}
\global\long\def\avg#1{\left< #1 \right>}
\let\underdot=\d{
\global\long\def\dd#1#2{\frac{d^{2}#1}{d#2^{2}}}
\global\long\def\pd#1#2{\frac{\partial#1}{\partial#2}}
\global\long\def\pdd#1#2{\frac{\partial^{2}#1}{\partial#2^{2}}}
\global\long\def\pdc#1#2#3{\left( \frac{\partial#1}{\partial#2}\right)_{#3}}
\global\long\def\op#1{\hat{\mathrm{#1}}}
\global\long\def\ket#1{\left| #1 \right>}
\global\long\def\bra#1{\left< #1 \right|}
\global\long\def\braket#1#2{\left< #1 \vphantom{#2}\right| \left. #2 \vphantom{#1}\right>}
\global\long\def\matrixel#1#2#3{\left< #1 \vphantom{#2#3}\right| #2 \left| #3 \vphantom{#1#2}\right>}
\global\long\def\av#1{\left\langle #1 \right\rangle }
 \global\long\def\com#1#2{\left[#1,#2\right]}
\global\long\def\acom#1#2{\left\{  #1,#2\right\}  }
\global\long\def\grad#1{\gv{\nabla} #1}
\let\divsymb=\div 
\global\long\def\div#1{\gv{\nabla} \cdot#1}
\global\long\def\curl#1{\gv{\nabla} \times#1}
\let\baraccent=\={
%
%
%

\begin{abstract}
\end{abstract}

\begin{frontmatter}

\title{Quantum simulation of zero temperature quantum phases and incompressible states of light via non-Markovian reservoir engineering techniques}


\author[paris]{Jos\'e Lebreuilly}
\ead{jose.lebreuilly@lpa.ens.fr}

\author[tn]{Iacopo Carusotto\fnref{postaladdress}}
\ead{carusott@science.unitn.it}

\address[paris]{Laboratoire Pierre Aigrain, D\'epartement de Physique de l'\'Ecole Normale Sup\'erieure, 24 rue Lhomond, 75231 Paris, France}
\address[tn]{INO-CNR BEC Center and Dipartimento di Fisica, Universit\`a di Trento, via Sommarive 14, I-38123 Povo, Italy}
\begin{abstract}
We review recent theoretical developments on the stabilization of strongly correlated quantum fluids of light in driven-dissipative photonic devices through novel non-Markovian reservoir engineering techniques. This approach allows to compensate losses and refill selectively the photonic population so to sustain a desired steady-state. It relies in particular on the use of a frequency-dependent incoherent pump which can be implemented, e.g., via embedded two-level systems maintained at a strong inversion of population. As specific applications of these methods, we discuss the generation of Mott Insulator (MI) and Fractional Quantum Hall (FQH) states of light.  As a first step, we present the case of a narrowband emission spectrum and show how this allows for the stabilization of MI and FQH states under the condition that the photonic states are relatively flat in energy. As soon as the photonic bandbwidth becomes comparable to the emission linewidth, important non-equilibrium signatures and entropy generation appear. As a second step, we review a more advanced configuration based on reservoirs with a broadband frequency distribution, and we highlight the potential of this configuration for the quantum simulation of equilibrium quantum phases at zero temperature with tunable chemical potential. As a proof of principle we establish the applicability of our scheme to the Bose-Hubbard model by confirming the presence of a perfect agreement with the ground-state predictions both in the Mott Insulating and superfluid regions, and more generally in all parts of the parameter space. Future prospects towards the quantum simulation of more complex configurations are finally outlined, along with a discussion of our scheme as a concrete realization of quantum annealing.
\end{abstract}


\end{frontmatter}

\section{Introduction}

Over the last few decades, a growing attention has been devoted to the study of many-body physics in the so-called {\em quantum fluids of light}~\cite{Carusotto_rev}: under a suitable confinement, photons acquire a finite effective mass and the optical nonlinearity of the medium can mediate interactions between photons. These advances were made possible through the development of several new experimental platforms such as exciton-polaritons in semiconducting microcavities\cite{Deng_rev}, superconducting circuits coupled to Josephson junctions \cite{circuits}, ultrastrong nonlinearities of coherently dressed atoms in a Rydberg-EIT configuration~\cite{Firstenberg_rev}. 

In addition to their own scientific interest for many-body physics~\cite{Carusotto_rev,Le_Hur_rev,Noh_rev,Hartmann_rev}, these photonic platforms hold a strong promise in view of quantum simulations~\cite{Houck_rev}: on one hand, their flexibility allows to engineer a wide range of Hamiltonian dynamics; on the other hand, their intrinsically driven-dissipative nature opens new routes towards the optical generation and the manipulation of the desired quantum state and offers the possibility of a direct read-out of the quantum state via the emitted light. Thanks to these remarkable features one can anticipate substantial advantages over more traditional platforms such as cold atomic gases or electronic systems.

The field of many-body physics with light experienced a sudden development in the late 2000's with a series of milestone results in gases of exciton-polaritons in semiconductor microcavities such as the demonstration of Bose-Einstein condensation  at both cryogenic~\cite{Kasprzak_BEC} and room~\cite{Kena_BEC} temperatures, along with superfluidity~\cite{Amo_SF} and associated quantum hydrodynamic effects such as solitons~\cite{Amo_soliton} and acoustic black holes~\cite{Black_holes}. Almost at the same time, a pioneering theoretical proposal by Haldane and Raghu \cite{Haldane_Ragu_Topo} gave rise to the new field of {\em topological photonics}~\cite{Topo_phot_rev}: in suitable configurations, neutral particles such as photons can be made to experience a synthetic gauge field \cite{artificial_gauge_1,artificial_gauge_2} in very much the same way as charged particles like electrons do in the presence of a real magnetic field. In this way, topologically protected edge states were observed in photonic lattices in the microwave~\cite{Wang_edge_MW} and soon later in the visible~\cite{Hafezi_edge,Rechtsman_edge} domains. Related studies in continuous-space geometries have led to the recent observation of Landau levels for photons in single twisted resonators~\cite{Schine_Landau}.

The present challenge is to access a regime of strong photon-photon interactions where strong correlations appear in the gas of photons and complex quantum many-body states can be studied~\cite{Carusotto_rev,Le_Hur_rev,Noh_rev,Hartmann_rev,Houck_rev}. While a regime of impenetrable photons via photon blockade  has already been observed in many experimental contexts at a single cavity/resonator level \cite{Birnbaum,Lang,Atac,Faraon,Peyronel,Jia_dot}, first experimental reports of many-mode dynamics of strongly interacting photons in extended geometries have started appearing, from strong antibunching and photon bound states in light propagating through coherently dressed atomic gases in the Rydberg-EIT configuration~\cite{Firstenberg_rev,Peyronel}, to a three-site lattice model for impenetrable photons in the presence of a strong synthetic magnetic field~\cite{Roushan_chiral_GS}. These two latter works, along with the recent observation of a driven-dissipative phase transition in a much larger circuit-QED one-dimensional chain featuring reasonably high interactions \cite{Fitzpatrick}, suggest that quantum simulation of incompressible quantum phases of light should be soon realized, including Mott insulators \cite{Fisher}, where interaction blockade prevents the onset of long range order and superfluidity, and fractional quantum Hall liquids \cite{Laughlin}, where topology and interactions contribute altogether to deeply modify the collective fluid properties. Very recently, first experimental evidence of a Mott insulator of light was reported at several conferences~\cite{Jon_private}.

While the first theoretical studies of strongly correlated fluids of light~\cite{Hartmann,Greentree,Angelakis,Koch_JCH} focused on equilibrium or quasi-equilibrium cases where photon losses and pumping can be neglected, in almost every realistic setup neither the particle number nor the energy are conserved quantities. Photon losses then play a crucial role and impose the necessity of some external pumping to continuously replenish the gas. As a result, even in the presence of a continuous pumping, the long-time state of the gas does not necessarily correspond to a thermal equilibrium condition, but rather to a {\em non-equilibrium steady-state}.

A direct strategy to circumvent this difficulty is of course to generate and manipulate the photon gas on a short time scale compared to the losses. This strategy was originally investigated in~\cite{Henriet,Tomadin1}, who proposed to use $\pi$-pulses to initialize a Mott insulator state, and very recently extended to fractional quantum Hall liquids in~\cite{Dutta}. Experimentally, this strategy was successfully implemented in~\cite{Roushan_chiral_GS} to study the few-body physics in small arrays. While its strong potential for relatively small systems is now established, strong difficulties can be expected to arise for large systems. In this case, adiabaticity constraints can in fact be much more stringent and the quantum phase transitions such as the Mott insulator to superfluid one are subject to critical slowing down phenomena and the consequent generation of domains in the ordered phase. Compared to cold atoms, the wider range of available manipulation schemes for photons is likely over-compensated by the detrimental effect of the much faster losses.

Simultaneously to the efforts to improve the photon lifetime and suppress the dissipative effects, alternative strategies that try to exploit the interplay between photon  losses and an external drive as a new feature of photonic systems have started being investigated. A first proposal in this direction has consisted in applying a continuous-wave coherent pump to the system~\cite{Carusotto_fermion,Tomadin,Umucalilar_FQH_coherent}.
On one hand, this approach is very appealing in virtue of its experimental and conceptual simplicity, and has been experimentally shown to display a very rich many-body physics of a non-equilibrium type in both continuum~\cite{Amo_SF,Amo_soliton,Baas_bistability} and lattice~\cite{Amo_bistability,Le_Boite_MF,Biondi,Foss-Feig} versions. In the context of strongly interacting photons, coherent pumping schemes have been proposed as a powerful tool to spectroscopically generate, manipulate, and probe strongly correlated states~\cite{Carusotto_fermion,Umucalilar_FQH_detection}. On the other hand, since a coherent drive scheme breaks the $U(1)$ symmetry related to the gauge transformation $a_i \to e^{i\phi}a_i$ on the various photonic annihilation operators $a_i$, one can anticipate that it is not ideal to guide a strongly interacting many-body system close to an incompressible quantum phase, as such a kind of state possesses by definition a well-defined total photon number and is thus invariant under the above-mentioned gauge transform. Likewise, the reversible nature of the photon absorption and emission processes under a coherent pump are typically responsible for additional undesired particle number fluctuations. For large system sizes, a very fast decrease of the overlap with the desired many-body eigenstate accompanied by a drastically different collective behaviour are thus expected.

In order to overcome these difficulties, autonomous stabilization schemes that allow to dynamically refill the many-body state after any undesired dissipative event have started attracting a growing interest.  In strict analogy with the idea of quantum error correction \cite{Kapit_error}, these setups are able to detect and automatically repair at a fast rate the defects (such as elementary excitations) generated within the quantum fluid due to the unavoidable presence of the external environment. This research direction sparkled with a pioneering work~\cite{Kapit} in the circuit-QED context regarding the stabilization of Fractional Quantum Hall states of light, and was further developed with our proposal for the generation of a photonic Mott Insulator \cite{Lebreuilly_2016}. The subsequent investigation of theses strategies over the last year by many studies \cite{Ma_Simon,Alberto_BH,Lebreuilly_square,Umucalilar_FQH_incoherent,Hartmann_non-Markovian}, along with the first announces of their experimental implementation and the observation of a Mott insulator state of light~\cite{Jon_private} motivate us to think that these methods will play a centrale role in the future generation of experiments. 

In this review, we summarize the last developments on the generation of strongly correlated states of light through  such autonomous stabilization schemes, with a particular focus on our own scientific production. We suggest to refer to this ensemble of techniques under the term of \textit{non-Markovian reservoir engineering}, since all the related proposals \cite{Kapit,Lebreuilly_2016,Ma_Simon,Alberto_BH,Lebreuilly_square,Umucalilar_FQH_incoherent,Hartmann_non-Markovian} are mainly exploiting the non-trivial spectral properties of a tailored artificial environment so to drive the photon gas into the desired quantum many-body state. In particular, the scheme involved in \cite{Lebreuilly_2016} relies on the inclusion of population-inverted emitters inside the photonic system hosting the fluid of light, so to implement an incoherent but frequency-dependent pumping. The simplicity of this scheme is amenable to experimental realization in a variety of different platforms, and a further stabilization of the many-body ground state is expected to be possible by using more advanced non-Markovian reservoir engineering protocols \cite{Lebreuilly_square}.

This article is organized as follows. In Sec.\ref{sec:reservoir_engineering} we review the theoretical framework of non-Markovian reservoirs coupled to a photonic system. In Sec.\ref{sec:narrow_band}, we discuss the case of Lorentzian-shaped, narrowband emitters: after a short presentation of single-site physics, we discuss in detail the generation of Mott insulator states, their peculiar non-equilibrium phase transitions towards superfluid states, and the generation of the simplest fractional quantum Hall states. In the following Sec.\ref{sec:broadband}, we extend our discussion to the case of more complex reservoirs with broadband spectra implementing respectively frequency-dependent photonic pump and loss processes, and we show how this last scheme allows for the quantum simulation of the ground-state of many-body Hamiltonians even in the regime of strong quantum correlations. In particular, the possibility of studying the zero temperature physics of incompressible fluids of light as well as the equilibrium Mott insulator-to-superfluid quantum phase transition are highlighted. Conclusions and future directions are finally outlined in Sec.\ref{sec:conclusions}.  

A relevant part of the works reviewed in this article constitutes the core of José Lebreuilly’s PhD thesis at the University of Trento~\cite{Lebreuilly_PhD} and was carried out in close collaboration with Alberto Biella, Florent Storme, Davide Rossini, Rifat Onur Umucal{\i}lar, Michiel Wouters, Rosario Fazio and Cristiano Ciuti. 

\section{Non-Markovian reservoir engineering}
\label{sec:reservoir_engineering}

\begin{figure}[t]
\centering
\includegraphics[width=0.7\textwidth,clip]{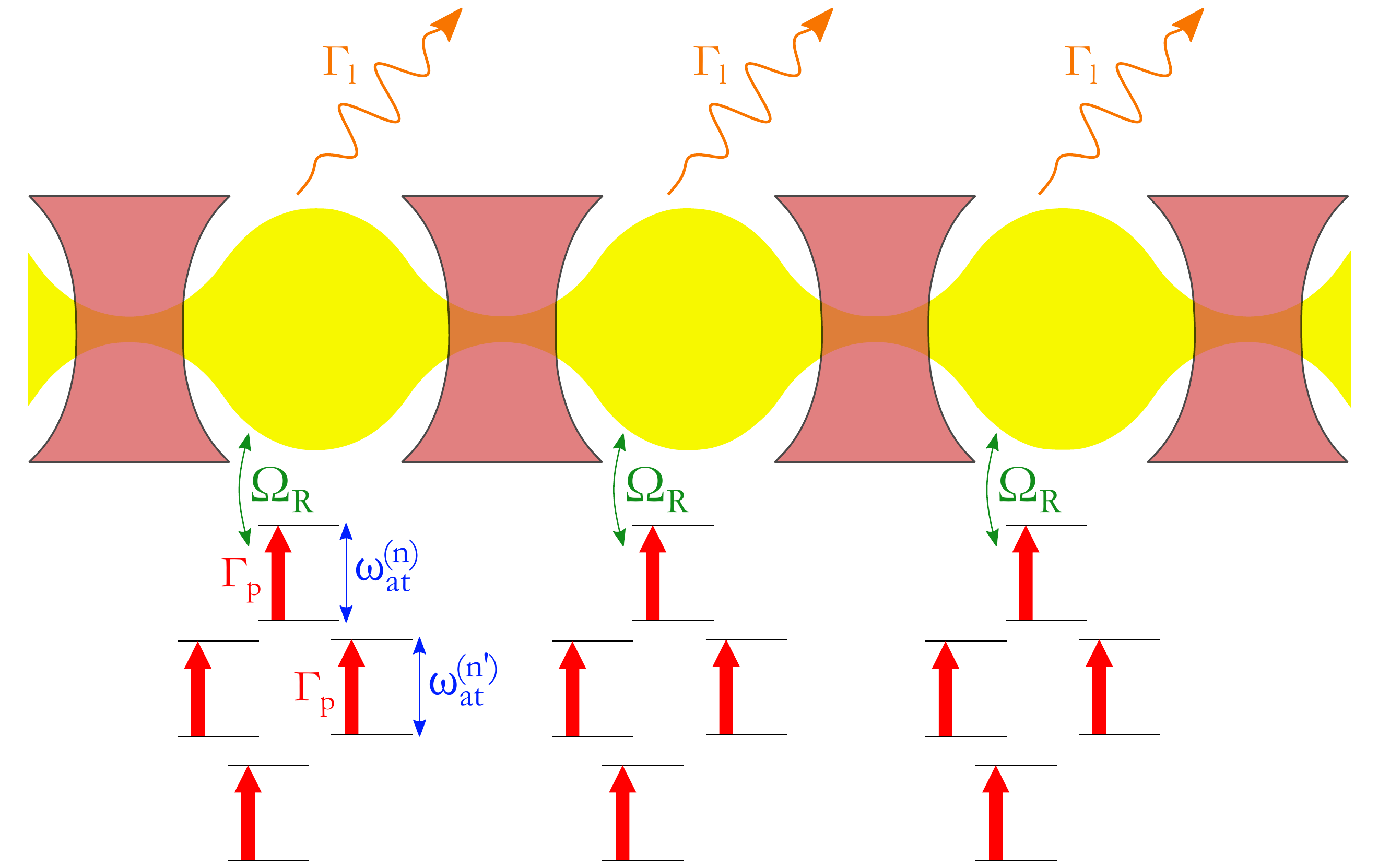} 
\hspace{1cm}
\caption{Non-Markovian reservoir engineering scheme for nonlinear cavity arrays or circuit QED. A set of two-level systems (atoms or qubits) with transition frequencies $\omega_{\rm{at}}^{(n)}$ is Rabi coupled to each lattice site (cavity or resonator) with a coupling strength $\Omega_{R}$. Each two-level system is fastly pumped toward its excited state at a strong rate $\Gamma_{\rm{p}}\gg\Omega_{R}$ in such a way to induce a strong inversion of population. Due to the interplay between the pumping and the Rabi coupling, the two-level systems are able to inject new photons inside the system in an incoherent (i.e., $U(1)$ symmetry preserving) but frequency-dependent manner, and photonic reabsorption processes are strongly suppressed. 
\label{fig:scheme}}
\end{figure}

We begin our presentation with the discussion of a concrete implementation of a non-Markovian reservoir in a driven-dissipative photonic device. While the various proposals \cite{Kapit,Lebreuilly_2016,Ma_Simon} suggested different ways to realize such reservoir, it can be shown that they are equivalent in the sense that they end up providing the same effective driven-dissipative photonic dynamics. Thus, in order to keep the discussion the most transparent, we will focus on the specific implementation scheme introduced in~\cite{Lebreuilly_2016} involving embedded two-level emitters with an inversion of population. We will consider as a first step a specific lattice model but generalization of this approach to other configurations is straightforward. Recent such developments with the purpose of stabilizing Fractional Quantum Hall states of light in single twisted resonators cavities in a continuous-space geometry are reviewed in Sec.~\ref{subsec:FQH}.

\subsection{The physical system}
\label{subsec:microscopic_model}
We consider a driven-dissipative Bose-Hubbard model for photons in an array of $L$ coupled nonlinear resonators of uniform bare frequency $\omega_{\rm{cav}}$. The experimental platform can be either be a superconducting circuit or a cavity QED one. In units such that $\hbar=1$, the Hamiltonian of the isolated photonic system takes the form~\cite{Carusotto_rev,Le_Hur_rev,Noh_rev,Hartmann_rev,Houck_rev}:
\begin{equation}
H_{\rm{ph}}=\sum_{i=1}^L\left[\omega_{\rm{cav}}a_{i}^{\dagger}a_{i}+\frac{U}{2}a_{i}^{\dagger}a_{i}^{\dagger}a_{i}a_{i}\right]{-}\sum_{\avg{i,j}}\left[ Ja_{i}^{\dagger}a_{j}+hc\right].
\end{equation}
where $a_{i}$ ($a_i^\dagger$) are bosonic annihilation (creation) operators for photons in the $i$-th resonator. The resonators are coupled via tunneling processes with amplitude $J$, and the lattice geometry and dimensionality are left unspecified at this stage for generality purposes. Each resonator is assumed to contain a Kerr nonlinear medium, which induces effective on-site repulsive interactions between photons with a strength $U$ proportional to the Kerr nonlinearity $\chi^{(3)}$. This nonlinearity can be obtained, e.g., by strongly coupling the photonic field in each resonator to a far-detuned two-level system. Unavoidable dissipative phenomena are responsible for photonic loss processes, which occur at a rate $\Gamma_{\rm{l}}$ and, unless differently specified, are modelled as fully Markovian, i.e., frequency independent.

We now present the non-Markovian reservoir engineering techniques which we will use in order to compensate the presence of losses and to guide the photonic population close toward a strongly correlated quantum phase. Instead of a coherent pumping, we consider a configuration, sketched in Fig.~\ref{fig:scheme}, where a set of $N_{\rm{at}}$ two-level emitters (differing from the one responsible for the Kerr effect) with different bare transition frequencies $\omega_{\rm{at}}^{(n)}$ is present in each resonator. Each two-level system, which can be, e.g., an atom or a qubit depending on the chosen experimental platform, is coupled to the resonator with a Rabi frequency $\Omega_{R}$ and is assumed to be strongly incoherently pumped toward its excited state at a rate $\Gamma_{\rm{p}}\gg\Omega_{\rm{R}}$ in such a way to maintain a strong inversion of population. In circuit QED setups, a strategy to implement this pumping was introduced in a more recent proposal \cite{Ma_Simon} relying on the coupling of a transmon qubit to a very lossy resonator tuned on the transition from the first to the second qubit excited state. In the optical regime such atomic pumping is relatively standard and can be obtained in analogy with a laser by mean of an hidden third atomic level \cite{Scully_book}. As we will see in the next section, the interplay between the inversion of population and the Rabi coupling to the photonic resonator mode makes the emitters to behave as a non-Markovian reservoir responsible for a frequency-dependent injection of photons into the system. Each emitter will provide an incoherent frequency-dependent photonic emission centered at the transition frequency $\omega_{\rm{at}}^{(n)}$, so that the total emission spectrum $\mathcal{S}_{\rm{em}}(\omega)$ results from the sum of the contributions of individual emitters and can display a complex $\omega$-dependence.

This physical expectation can be put on solid grounds in terms of a microscopic model describing the coupling of the emitters to light. The free evolution of the emitters and their coupling to the resonators are described by the following Hamiltonian terms,
\begin{eqnarray}
H_{\rm{at}}&=&\sum_{i=1}^L\sum_{n=1}^{N_{\rm{at}}}\omega_{\rm{at}}^{(n)}\sigma_{i}^{+(n)}\sigma_{i}^{-(n)} \\ 
H_{I}&=&\Omega_{R}\sum_{i=1}^L\sum_{n=1}^{N_{\rm{at}}}\left[a_{i}^{\dagger}\sigma_{i}^{-(n)}+a_{i}\sigma_{i}^{+(n)})\right],
\end{eqnarray}
the $\sigma_{i}^{\pm(n)}$ operators being the usual raising and lowering operators for the two-level $n$-th emitter in the $i$-th resonator. The dissipative dynamics under the effect of the pumping and decay processes can be described in terms of a master equation for the density matrix $\rho_{\rm{tot}}$ including both emitter and photonic degrees of freedrom, 
\begin{equation}
\partial_{t}\rho_{\rm{tot}}=\frac{1}{i}\com{H_{\rm{ph}}+H_{\rm{at}}+H_{I}}{\rho_{\rm{tot}}}+\mathcal{L}(\rho_{\rm{tot}}),\label{eq:evinitio}
\end{equation}
where the different dissipative processes are summarized in the Lindblad super-operator $\mathcal{L}=\mathcal{L}_{\rm{l}}+\mathcal{L}_{\rm{p}}$, with 
\begin{eqnarray}
\mathcal{L}_{\rm{l}}(\rho_{\rm{tot}})&=&\frac{\Gamma_{\rm{l}}}{2}\sum_{i=1}^{L}\left[2a_{i}\rho_{\rm{tot}} a_{i}^{\dagger}-a_{i}^{\dagger}a_{i}\rho_{\rm{tot}}-\rho_{\rm{tot}} a_{i}^{\dagger}a_{i}\right]\\
\mathcal{L}_{\rm{p}}(\rho_{\rm{tot}})  &=&  \frac{\Gamma_{\rm{p}}}{2}\sum_{i=1}^{L}\sum_{n=1}^{N_{\rm{at}}}\left[2\sigma_{i}^{+(n)}\rho_{\rm{tot}}\sigma_{i}^{-(n)}-\sigma_{i}^{-(n)}\sigma_{i}^{+(n)}\rho_{\rm{tot}}-\rho_{\rm{tot}}\sigma_{i}^{-(n)}\sigma_{i}^{+(n)}\right],
\end{eqnarray}
describing respectively the photon losses and the emitter pumping.

Before moving to the description of the effective non-Markovian photonic dynamics emerging from our microscopical model, for the sake of completeness it is worth quickly summarizing the main points of the alternative proposal in~\cite{Kapit}: instead of a population inverted two-level system, it suggests to employ a very lossy resonator of frequency $\omega_{\rm{at}}$   in order to provide the incoherent frequency-dependent pumping. As it is not possible to invert the population of an harmonic oscillator, it proposes to parametrically modulate the coupling  between the resonator and the photonic lattice at a frequency $2\omega_{\rm{at}}$ in such a way that the transition consisting in adding (resp. removing) a photon both in the resonator and the lattice becomes resonant: in this way, the lossy resonator ends up playing the same role as the emitter in our incoherent pumping model.

\subsection{An effective photonic non-Markovian description}

In the regime we considered, the concentration of the two-level emitters population into the excited state allows to write a closed master equation for the photonic density matrix $\rho_{ph}=Tr_{\rm{at}}\rho$ where the emitters degrees of freedom have been traced out. Our derivation, which is performed by mean of projective methods \cite{Breuer,Quantum_Noise}, can be found in \cite{Lebreuilly_2016}. The resulting photonic master equation reads 
\begin{equation}
\partial_{t}\rho =  -i\left[H_{\rm{ph}},\rho(t)\right]+\mathcal{L}_{\rm{l}}(\rho) + \mathcal{L}_{\rm{em}}(\rho),
\label{eq:photon_only}
\end{equation}
with 
\begin{eqnarray}
\mathcal{L}_{\rm{l}} (\rho)& = & \frac{\Gamma_{\rm{l}}}{2}\sum_{i=1}^{L}\left[2a_{i}\rho a_{i}^{\dagger}-a_{i}^{\dagger}a_{i}\rho-\rho a_{i}^{\dagger}a_{i}\right],\label{eq:loss}\\
\mathcal{L}_{\rm{em}}(\rho) & = & \frac{\Gamma_{\rm{em}}^0}{2}\sum_{i=1}^{L}\left[\tilde{a}_{i}^{\dagger}\rho a_{i}+a_{i}^{\dagger}\rho\tilde{a}_{i}-a_{i}\tilde{a}_{i}^{\dagger}\rho-\rho\tilde{a}_{i}a_{i}^{\dagger}\right]. \label{eq:gainmarkov}
\end{eqnarray}
describing photonic losses and emission processes, respectively.
While the loss term has a standard Lindblad form with the corresponding rate $\Gamma_{\rm{l}}$, the emission term keeps some memory of the emitters dynamics as it involves modified lowering and raising operators 
\begin{eqnarray}
\frac{\Gamma^{0}_{\rm{em}}}{2}\tilde{a}_{i} &=& \!\int_{0}^{\infty}d\tau\, \Gamma_{\rm{em}}(\tau)a_{i}(-\tau),\label{eq:atilde} \\
\tilde{a}_{i}^{\dagger}&=&\left[\tilde{a}_{i}\right]^{\dagger} \label{eq:adagtilde}
\end{eqnarray}
which contain the photonic hamiltonian dynamics during the emitter re-pumping $a_{i}(\tau)=e^{iH_{\rm{ph}}\tau}\,a_{i}\,e^{-iH_{\rm{ph}}\tau}$. The memory kernel $\Gamma_{\rm{em}}(\tau)=\theta(\tau)\int \frac{d\omega}{2\pi}\mathcal{S}_{\rm{em}}(\omega)e^{-i\omega\tau}$
involved in the expression of these modified operators takes into account the presence of a frequency-dependent emission power spectrum $\mathcal{S}_{\rm{em}}(\omega)$, which encapsulates the spectral properties of the reservoir of two-level emitters, and can be expressed in function of the parameters of the microscropic model of Sec.~\ref{subsec:microscopic_model} as 
 \begin{equation}
\label{eq:power-spectrum}
\mathcal{S}_{\rm{em}}(\omega)=\Gamma_{\rm{em}}^{\rm{at}}\int d\tilde{\omega}\mathcal{D}_{\rm{at}}(\tilde{\omega})\frac{(\Gamma_{\rm{p}}/2)^2}{(\omega-\tilde{\omega})^2+(\Gamma_{\rm{p}}/2)^2},
\end{equation}
where $\Gamma_{\rm{em}}^{\rm{at}}=4\Omega_{R}^2/\Gamma_{\rm{p}}$, and $\mathcal{D}_{\rm{at}}(\tilde{\omega})$ is the distribution of the two-level emitters transition frequencies $\omega_{\rm{at}}^{(n)}$. The constant $\Gamma^{0}_{\rm{em}}$ of Eq.~(\ref{eq:atilde}) has the dimension of a frequency, and is defined as the maximal value reached by the emission power spectrum $\Gamma_{\rm{em}}^0=\text{Max}_{\omega}\left[\mathcal{S}_{\rm{em}}(\omega)\right]$. The expression Eq.~(\ref{eq:power-spectrum}) for $\mathcal{S}_{\rm{em}}(\omega)$ can be understood intuitively as following: due to the spectral broadening induced by the pumping into the excited state, each emitter displays a Lorentzian spectral density of linewidth $\Gamma_{\rm{p}}$ centered around its transition frequency $\omega_{\rm{at}}^{(n)}$. The integration over the contribution of all emitters across their frequency distribution $\mathcal{D}_{\rm{at}}(\omega)$ then yields Eq.~(\ref{eq:power-spectrum}).

More physical insight on the quantity $\mathcal{S}_{\rm{em}}(\omega)$ can also be obtained by looking at the matrix elements of the operators (\ref{eq:atilde}) and (\ref{eq:adagtilde}) in the eigenbasis photonic hamiltonian $H_{\rm{ph}}$
\begin{equation}
\label{eq:special-operators-frequency}
\bra{f}\tilde{a}_i\ket{f'} =\frac{2}{\Gamma_{\rm{em}}^0} \Gamma_{\rm{em}}(\omega_{f'f}) \bra{f}a_i\ket {f'}:
\end{equation}
here  $\ket f$ (resp. $\ket{f'}$) are two many-body eigenstates of $H_{\rm{ph}}$ with $N$ (resp. $N+1)$ photons and energy difference $\omega_{f'f}=\omega_{f'}-\omega_{f}$, and 
\begin{equation}
\Gamma_{\rm{em}}(\omega) = \frac{1}{2} \mathcal{S}_{\rm{em}}(\omega)-i\delta_{l}(\omega)
\end{equation}
is the Fourier transform of the memory kernel $\Gamma_{\rm{em}}(\tau)$. The magnitude of the Lamb-shift $\delta_{l}(\omega)$ stemming from the imaginary part of $\Gamma_{\rm{em}}(\omega)$ is typically small as compared to the emission linewidth $\Gamma_{\rm{em}}(\omega)$ and will not bring important physical effects in the physics we will discuss. The role of the real part $\mathcal{S}_{\rm{em}}(\omega)/2$ is instead a crucial one and becomes physically transparent in the secular regime (i.e., $\mathcal{S}_{\rm{em}}(\omega),\Gamma_{\rm{l}}\ll U,J$), where it gives rise to a frequency-dependent transition rate for the $\ket f\longrightarrow \ket{f'}$ transition,
\begin{equation}
\mathcal{T}(f\to f')=\mathcal{S}_{\rm{em}}(\omega_{f',f})\sum_i|\bra{f'} a_i^\dagger \ket{f}|^2.
\end{equation}
This expression for $\mathcal{T}(f\to f')$, which is the product of an internal contribution to the system describing the wave-function overlap between different many-body eigenstates up to the local addition/suppression of a single photon, and an external contribution involving only the reservoir spectral properties, could have alternatively been derived in this same regime by mean of the Fermi Golden rule. Such rate is to be compared to the one of the reciprocal process induced by losses,
\begin{eqnarray}
\mathcal{T}(f'\to f)&=&\Gamma_{\rm{l}}\sum_i|\bra{f} a_i \ket{f'}|^2=\Gamma_{\rm{l}}\sum_i|\bra{f'} a_i^\dagger \ket{f}|^2:
\end{eqnarray}
Strikingly, one has that the ratio
\begin{equation}
\frac{\mathcal{T}(f\to f')}{\mathcal{T}(f'\to f)}=\frac{\mathcal{S}_{\rm{em}}(\omega_{f',f})}{\Gamma_{\rm{l}}}
\end{equation}
between these two rates only depends on the spectral properties of the various baths around the transition frequency $\omega_{f',f}$, and does not involve more refined information on the internal structure of the many-body eigenstates  $\ket f$ and $\ket{f'}$. As the two processes stem from the contact with different physical reservoirs, the frequency dependency of $\mathcal{T}(f\to f')$ and $\mathcal{T}(f'\to f)$ can in principle be engineered completely independently from each other (in particular, a scheme involving additional frequency-dependent losses will be discussed in Sec.~\ref{subsec:frequency-dependent_losses}): such tunability is at the core of the idea of non-Markovian reservoir engineering techniques, as this will allow to guide the population toward desired eigenstates of the many-body Hamiltonian $H_{\rm{ph}}$, e.g., by targeting some specific transition frequencies or by suppressing all transitions above a certain energy cutoff. The former case is characterized by all transition frequencies $\omega_{\rm{at}}^{(n)} $ taking a unique value $\omega_{\rm{at}}$ was considered e.g. in~\cite{Kapit,Lebreuilly_2016,Ma_Simon,Alberto_BH,Lebreuilly_square,Umucalilar_FQH_incoherent} and is reviewed in the next Sec.\ref{sec:narrow_band}. The latter case, 
introduced in~\cite{Lebreuilly_square} and reviewed in Sec.~\ref{sec:broadband}, is characterized by the transition frequencies of the emitters being uniformly distributed over the interval $[\omega_-,\omega_+]$ and allows for a more complete quantum simulation of equilibrium physics. Strikingly, even though the initial proposals in~\cite{Kapit,Lebreuilly_2016,Ma_Simon} rely on different microscopic models, they all end up recovering the same  photonic effective theory Eq.~(\ref{eq:photon_only}) involving non-Markovian dynamics. It is thus possible to regroup all this models in a single framework, and to provide unified and robust predictions about the properties of the steady-state.

\section{Strongly interacting photons under a narrowband incoherent emission and Markovian losses}
\label{sec:narrow_band}
In this section we review the physics of strongly interacting photons under a narrowband frequency-dependent incoherent pump. Such a configuration, which is the one involved in all the first proposals \cite{Kapit,Lebreuilly_2016,Ma_Simon}, was further investigated in \cite{Alberto_BH,Umucalilar_FQH_incoherent}. Within the framework of the non-Markovian reservoir engineering techniques presented in the previous section, a narrowband incoherent emission is straightforwardly obtained by setting all emitters transition frequencies $\omega_{\rm{at}}^{(n)}$ to the unique value
$\omega_{\rm{at}}$ (or alternatively by using only $N_{\rm{at}}=1$ emitter per resonator): this leads to the expression $\mathcal{D}_{\rm{at}}(\omega)=N_{\rm{at}}\delta(\omega-\omega_{\rm{at}})$ for the emitters frequency distribution and to a Lorentzian-shaped photonic emission spectrum:
\begin{equation}
\label{eq:Lorentzian_emission}
\mathcal{S}_{\rm{em}}(\omega)=\mathcal{S}_{\rm{em}}^{\rm{Narrow}}(\omega)\equiv\Gamma_{\rm{em}}^{0}\frac{(\Gamma_{\rm{p}}/2)^2}{(\omega-\omega_{\rm{at}})^2+(\Gamma_{\rm{p}}/2)^2}.
\end{equation}
The modified jump operators presented in Eqs.~(\ref{eq:atilde})-(\ref{eq:adagtilde}) can be accordingly reexpressed in the eigenbasis of $H_{\rm{ph}}$ as:
\begin{eqnarray}
\bra f\tilde{a_{i}}\ket{f'} &=&  \frac{\Gamma_{\rm{p}}/2}{-i(\omega_{f'f}-\omega_{\rm{at}})+\Gamma_{\rm{p}}/2}\bra{f} a_{i}\ket{f'},\label{eqz:anni_lorent}\\
\bra {f'}\tilde{a_{i}}^\dagger\ket{f} &=&  \bra f\tilde{a_{i}}\ket{f'}^*,\label{eq:crea_lorent}
\end{eqnarray}
so that the ratio between the corresponding transition rates reads
\begin{equation}
\frac{\mathcal{T}(f\to f')}{\mathcal{T}(f'\to f)}=\frac{\Gamma_{\rm{em}}^{0}}{\Gamma_{\rm{l}}}\frac{(\Gamma_{\rm{p}}/2)^2}{(\omega-\omega_{\rm{at}})^2+(\Gamma_{\rm{p}}/2)^2}.
\end{equation}
New photons are efficiently emitted in to the system only where the corresponding many-body transition is close to resonance with the emitters frequency $\omega_{\rm{at}}$. In the next two subsections we will briefly present how this scheme allows to generate and stabilize photonic Mott Insulators in a dissipative Bose-Hubbard model \cite{Lebreuilly_2016,Ma_Simon,Alberto_BH}, and then a fractional quantum Hall state of light in two different geometries \cite{Kapit,Umucalilar_FQH_incoherent}.

\subsection{Bose-Hubbard model: stabilization of strongly localized Mott Insulating states}
\label{subsec:BH_narrwband}
A Mott insulator state of light would be characterized among other criteria by all sites displaying an integer average occupation, along with suppressed fluctuations in the lattice total particle number~\cite{Fisher}. As a first step in the direction of realizing such a strongly correlated quantum phase, we will discuss the toy-model case of a single strongly nonlinear cavity and assess the feasibility of  using the frequency-dependent incoherent pumping scheme to prepare it in a Fock state with a well-defined photon number. These results will be the starting point to generalize our discussion to extended many-cavity systems. From a quantum optics perspective, a Mott insulator state is an extreme manifestation of the corpuscular nature of photons.

\subsubsection{Single cavity physics: Fock states stabilization}
\begin{figure}[t]
\centering
\includegraphics[scale=0.30,clip]{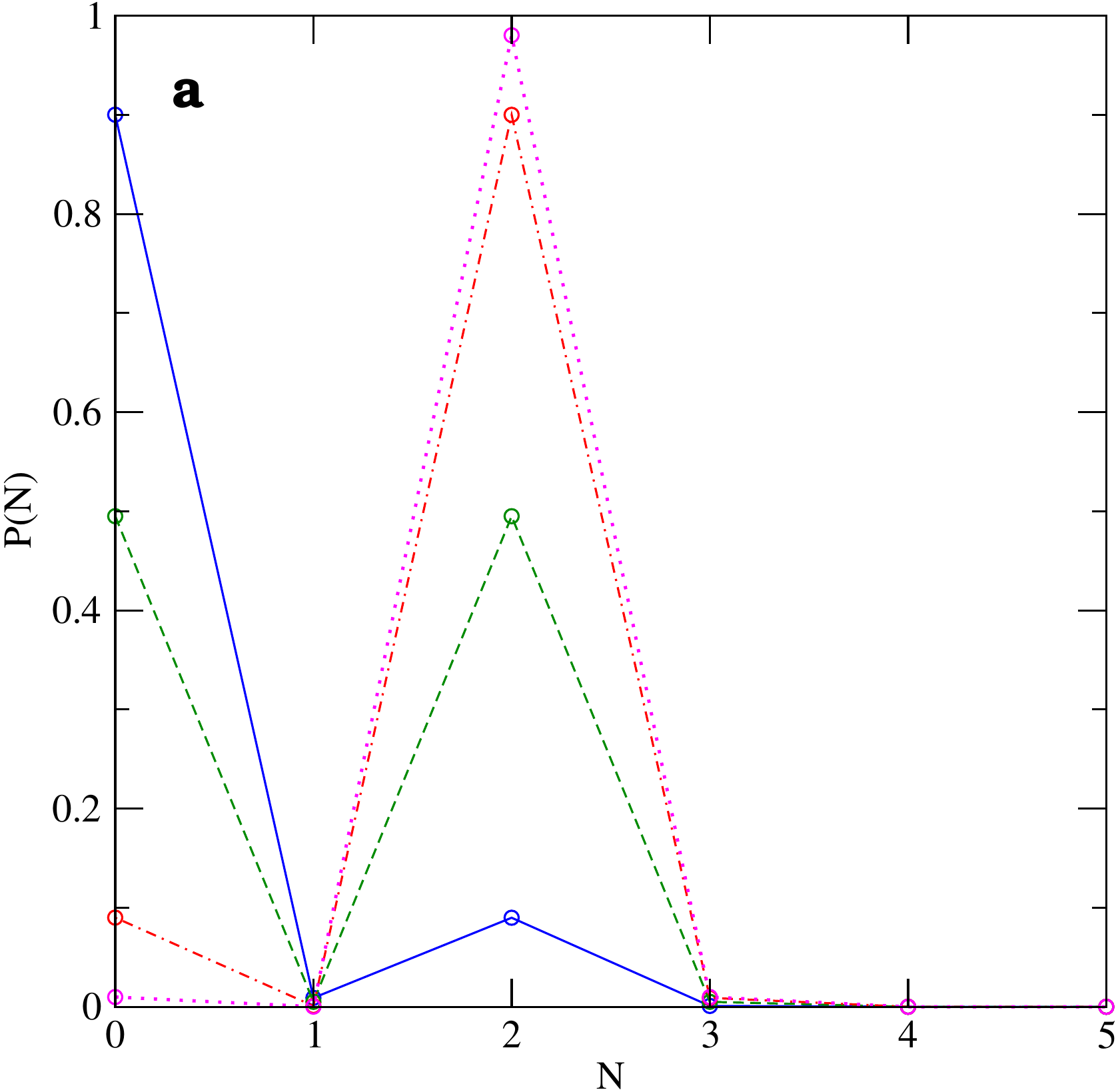} 
\hspace{1cm}\includegraphics[scale=0.3,clip]{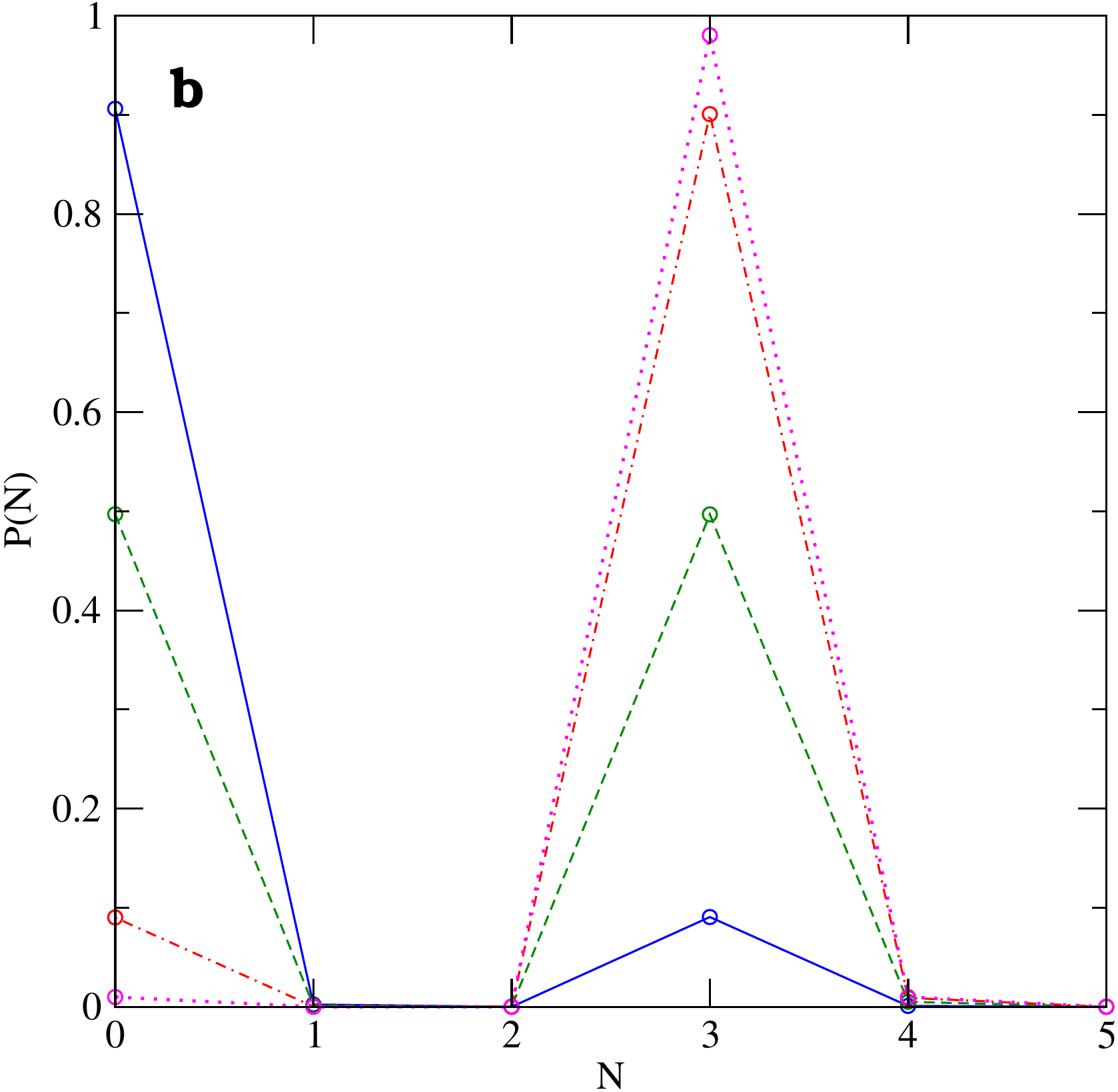}
\caption{Selective generation of a $N_0=2$ photon [panel a)]  and $N_0=3$ photon [panel b)] Fock state: Population $P(N)$ as a function of $N$. In each panel, the various lines correspond to the steady-state properties for different choices of pumping parameters, allowing to reach various levels of fidelity with the targeted $N_0$ Fock state. Figure adapted from \cite{Lebreuilly_2016}.
\label{fig:one_strongly}}
\end{figure}

As it was discussed in~\cite{Lebreuilly_2016}, the photonic effective non-Markovian master equation (\ref{eq:photon_only}) for a single cavity possesses an exact analytical solution for the steady-state density matrix $\rho_\infty=\sum_N P(N)\ket{N}\bra{N}$, where
\begin{equation}
\label{eq:single-cavity_solution}
P(N) = {\pi_{0}}\,\prod_{M=0}^{N-1}\frac{\mathcal{S}_{\rm{em}}(\omega_{M+1,M})}{\Gamma_{\rm{l}}}= \left(\frac{\Gamma_{\rm{em}}^0}{\Gamma_{\rm{l}}}\right)^{N}\pi_{0}\prod_{M=0}^{N-1}\frac{(\Gamma_{\rm{p}}/2)^{2}}{(\omega_{\rm{cav}}+UN-\omega_{\rm{at}})^{2}+(\Gamma_{\rm{p}}/2)^{2}}.
\end{equation}
is the probability of occupancy of the $N$-photon Fock state. The solution Eq.~(\ref{eq:single-cavity_solution}) stems from a simple and analytically exact detailed balance relation
\begin{equation}
0=\mathcal{T}(N+1\to N)P(N+1)-\mathcal{T}(N\to N+1)P(N)
\end{equation}
which emerges only for a $L=1$ site system, an does not apply anymore for larger lattices. In the weakly nonlinear regime $U\ll\Gamma_{\rm{p}}$, our analysis unveiled an exotic bistability mechanism  differing from the one typically studied in classical nonlinear optics~\cite{Scully_book} as well as from the ones discussed in earlier works on strongly correlated photons~\cite{Le_Boite_MF,Biondi}. 

Most remarkably, in the regime of a strong nonlinearity $U\gg\Gamma_{\rm{p}}$ and a strong photonic emission $\Gamma_{\rm{em}}^0/\Gamma_{\rm{l}}\to +\infty$, our narrowband non-Markovian scheme straightforwardly allows to stabilize photonic Fock states with a well-defined photon number $N_0$, as shown in Fig.~\ref{fig:one_strongly}. The selection of a specific number $N_0$, can be done through an appropriate choice of a detuning $\omega_{\rm{at}}=\omega_{\rm{cav}}+(N_0-1)U$ between the bare cavity frequency and the emitters transition frequency. This ability to obtain a strong selectivity for the photonic population and ultimately concentrate all the probability in the targeted Fock state (see purple dotted lines in both panels of  Fig.~\ref{fig:one_strongly}) is a direct consequence of our choice of an incoherent non-Markovian pump scheme, and could not have been obtained by mean of a standard coherent drive: indeed, in the typical photon blockade regime, even at very strong coherent drive intensities the population would at best be equally distributed between the zero- and the one-photon states. On the other hand, the fast repumping of the two-level emitters of the non-Markovian reservoirs into their excited state allows to fully suppress reabsorption processes and Rabi oscillations with the cavity mode, and thus leads to a fully irreversible emission, which is a necessary ingredient for the suppression of photonic number fluctuations.

While the case $N_0=1$ of a single photon Fock state is relatively trivial, and is reachable by mean of the state-of-the-art technology of circuit QED, we noticed that having a sizeable stationary population in the $N = N_0$ peak requires quite extreme values of the parameters for higher photon number, as they become exponentially high in $N_0$. The physics underlying these difficulties can be easily explained in terms of the asymmetry in the switching mechanisms leading from $N=0$ to $N=N_0$ and viceversa.

Fortunately, even in the situation where all the population is not concentrated in the desired Fock state (e.g., in the green dashed lines and blue solid lines of Fig.~\ref{fig:one_strongly}), one can prove that the $N=N_0>1$ state can be dynamically statibilized over a time $\tau\propto\frac{1}{\Gamma_{\rm{l}}}\frac{\Gamma_{\rm{em}}^0}{\Gamma_{\rm{l}}}$ much longer than the single photon lifetime $\frac{1}{\Gamma_{\rm{l}}}$, using parameters of the same order of magnitude as for the $1$-photon Fock state: indeed, unstabilize the $N=N_0$ state would require two successive photon loss events, which is very unlikely as the two-level emitters will refill the $N=N_0-1$ state with a much higher efficiency after the first photon loss process. Alternatively, remaining in the same range of technologically accessible pumping parameters, the $N=N_0$ can be fully stabilized (i.e., $P(N_0)\simeq 1$) by using $N_{\rm{at}}=N_0$ different species of emitters with transition frequencies $\omega_{\rm{at}}^{(n)}=\omega_{\rm{cav}}+(n-1)U$ tuned in such a way to protect all transitions $n\to n+1$ for $0\leq n\leq N_0-1$. This idea is further explored in Sec~\ref{sec:broadband} where we review, among other features, the single-cavity steady-state properties in presence of a tailored square-shaped emission spectrum allowing to cover all photonic transitions and thus fully stabilize arbitrary Fock states with an high efficiency.

\subsubsection{Lattice physics: Stabilization of strongly localized $n=1$ Mott states}
\label{subsubsec:flat_Mott}
Extending the photon-number selectivity idea to the many-cavity case, we now look for many-body states that resemble a Mott insulator state in the $U\gg\Gamma_{\rm{p}}$ strong interactions limit. As in the single cavity case discussed in the previous subsection, the strong pumping $\Gamma^0_{\rm{em}}\gg \Gamma_{\rm{l}}$ tends to favour a large occupations of sites, but is counteracted, for a zero detuning $\omega_{\rm{at}}=\omega_{\rm{cav}}$, by the effect of the nonlinearity $U\gg \Gamma_{\rm{p}}$ which set an upper bound to the occupation by forbidding the addition of a second photon on top of an already existing one. In the zero-tunneling $J=0$ case, we of course trivially recover the single cavity physics and our scheme predictably leads to the formation of a perfect Mott state with $n=1$ photon per site.  

We now move to the more complex  case of a weak but non-vanishing tunneling constant $J\ll\Gamma_{\rm{p}}\ll U$. Since $J\lll U$,  photons are still unable to overcome photon blockade by tunneling and quantum processes involving particle exchange are largely suppressed. In particular, the eigenstate of the photonic Hamiltonian $H_{\rm{ph}}$ corresponding to the Mott Insulator (with  $N=L$ photons)  is completely localized: photons are almost perfectly pinned on a single site. Another consequence of having a weak tunneling is that all transitions frequencies $\omega_{\rm{cav}}-\epsilon_k$ (with $-zJ<\epsilon_{k}<zJ$, $z$ being the number of lattice site nearest number) from hole excited states with momentum $k$ toward a completely saturated Mott-state have almost resonant values with the pump frequency $\omega_{\rm{at}}=\omega_{\rm{cav}}$ and fall within the emission frequency range (since $J\ll \Gamma_{\rm{p}}$). As a consequence, if one loses a photon starting from a completely filled Mott state with exactly one photon per site, the strong pump ($\Gamma^0_{\rm{em}}\gg \Gamma_{\rm{l}}$) will inject immediately a new photon and remove the corresponding hole excitation. One expects that in this regime the steady-state will still be a perfectly localized Mott-state with a well-defined $n=1$ number of photons per cavity.

\begin{figure}[t]
\centering
\includegraphics[scale=0.3,clip]{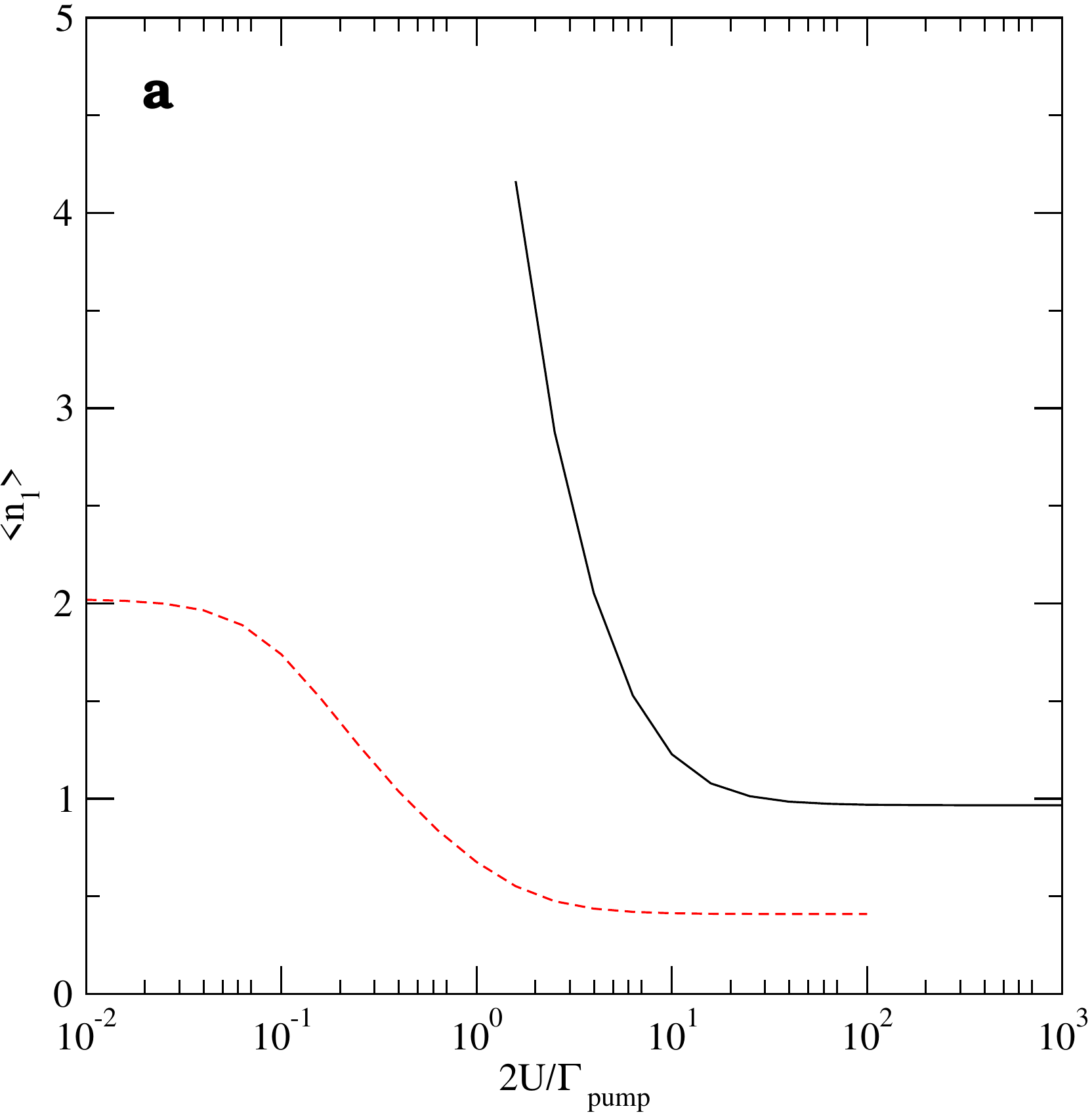} 
\includegraphics[scale=0.3,clip]{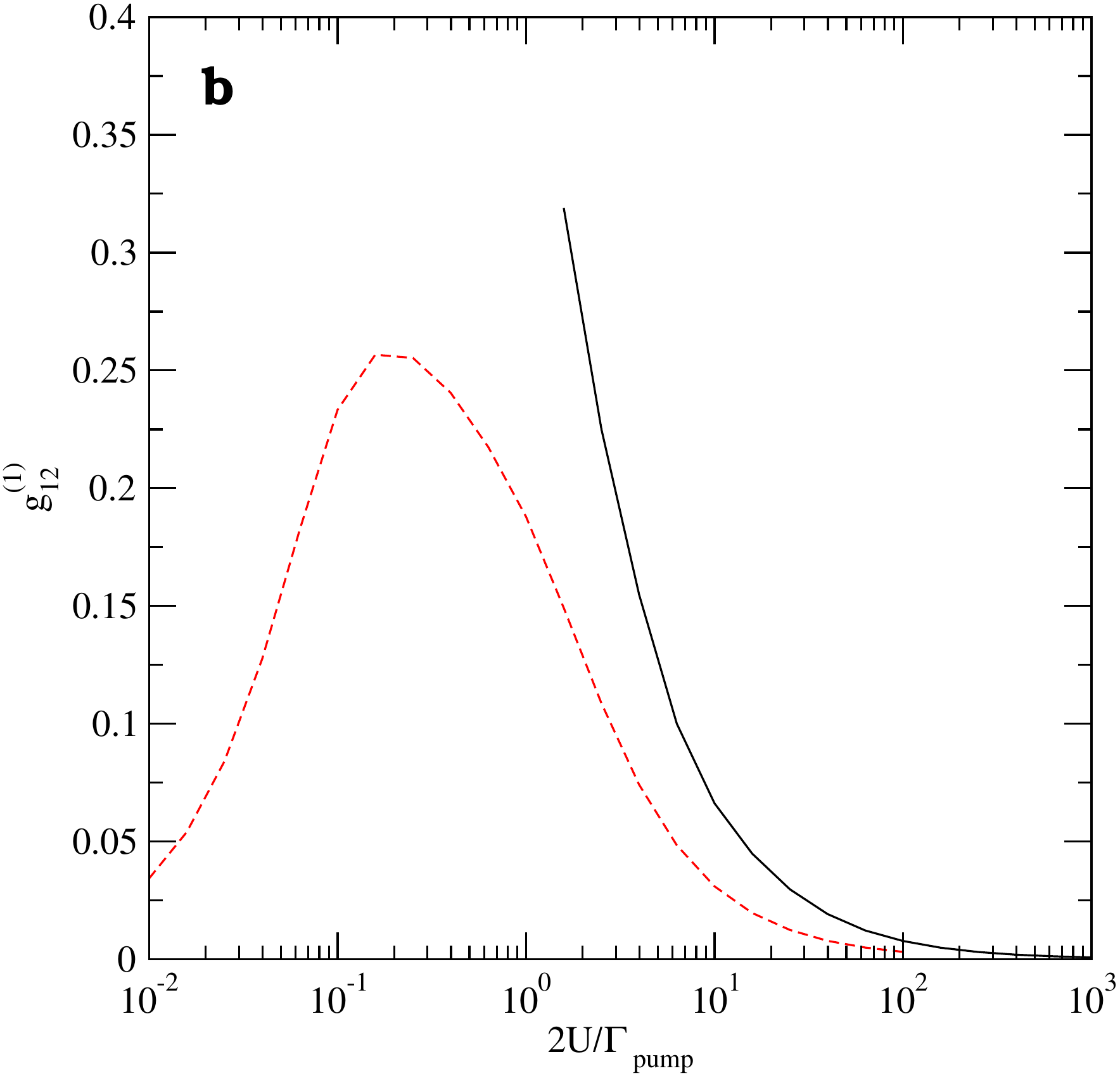} 
\includegraphics[scale=0.3,clip]{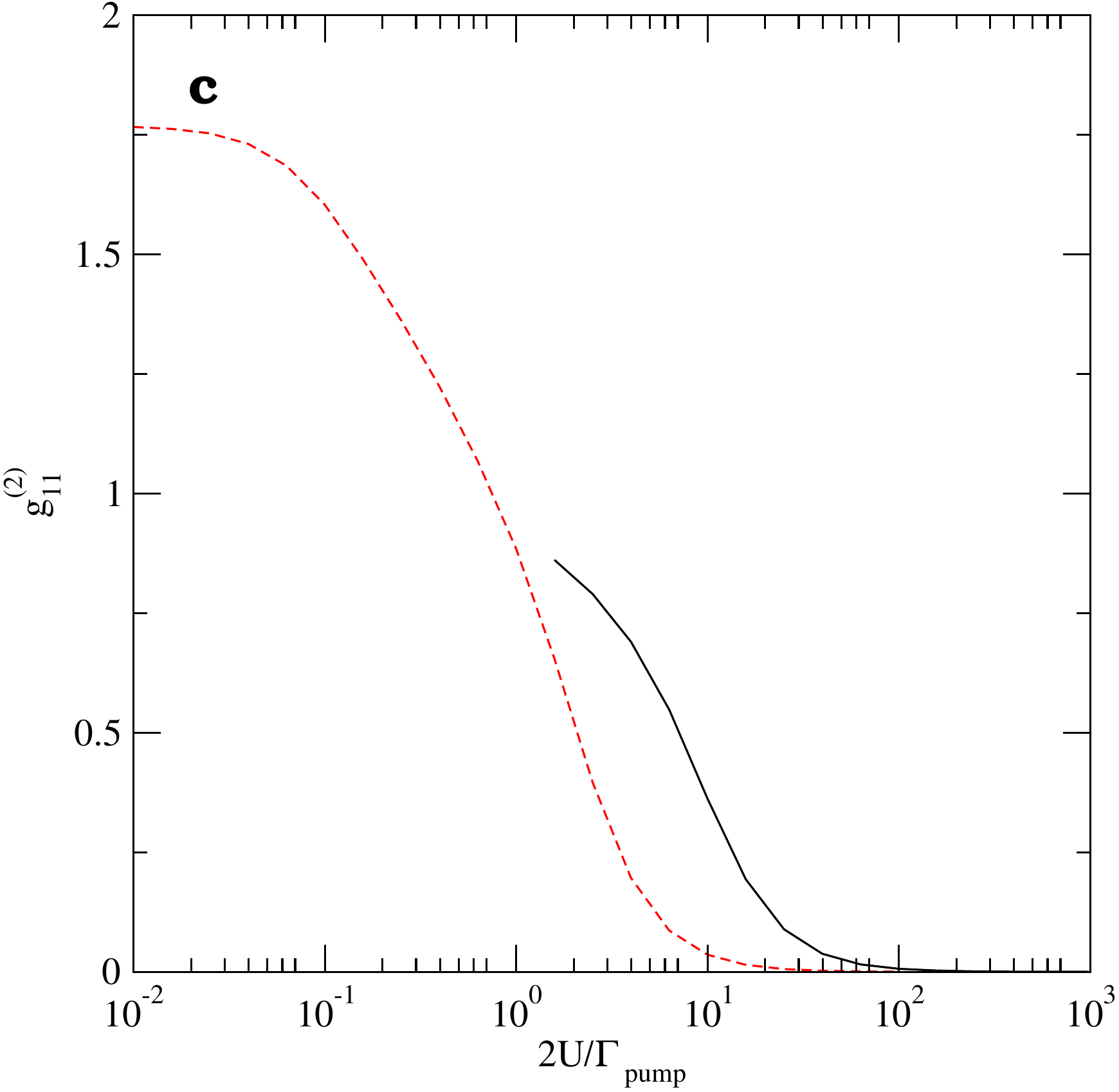}
\hspace{1cm}
\caption{Steady-state observables as a function of $2U/\Gamma_{\rm{p}}$ for a $L=2$ two-cavity system ($\Gamma_{\rm{p}}=\Gamma_{\rm{pump}}$ within our notation system): (a) average number of photons $n_{1}=\langle{a_{1}^{\dagger}a_{1}}\rangle$, (b) inter-site one-body correlation function $g_{1,2}^{(1)}={\langle a_{1}^{\dagger}a_{2}\rangle }/{\langle a_{1}^{\dagger}a_{1}\rangle }$, (c) one-site two-body correlation function $g_{1,1}^{(2)}=\langle{a_{1}^{\dagger}a_{1}^{\dagger}a_{1}a_{1}}\rangle=\langle{n_{1}(n_{1}-1)}\rangle$. The choice of parameters for the solid black line was done in such a way to favor the emergence of a $n=1$ Mott-like phase. Identical parameters were chosen for the red dashed line, at the exception of a weaker emission rate $\Gamma_{\rm{em}}^0$, in such a way to study the low-density phenomenology (not presented in this review).
Figure adapted from \cite{Lebreuilly_2016}.
\label{fig:mott}}
\end{figure}

This intuition is confirmed in Fig.~\ref{fig:mott}(a-c) (black solid lines) where we can see clear signatures of the desired Mott state with one particle per site in a two coupled cavity geometry : for an high emission rate $\Gamma_{\rm{em}}^0/\Gamma_{\rm{l}}\gg 1$ and a strong nonlinearity $U/\Gamma_{\rm{p}}\gg 1$, the steady-state average number of photons  [panel a)] and the probability of double occupancy [panel (b)] respectively tends to $1$ and $0$.  Finally the one-body coherence between two neighbouring sites also tends to $0$ [panel c)], confirming the photonic localization effect. This preliminary analysis on a very small $L=2$ lattice already allowed to make definite claims on the structure of the non-equilibrium phase diagram of our model and draw positive conclusions about the presence of a strongly localized Mott state~\cite{Lebreuilly_2016}. All these features were then confirmed by a more advanced numerical study \cite{Alberto_BH} performed in larger 1D chains up to $L=8$, which also found, as expected, a vanishing entropy $\mathcal{S}=-\text{Tr}\left[\rho_{\infty}\rm{ln}(\rho_{\infty})\right]$ and a full fidelity with the Hamiltonian ground-state $\rho_{\infty}=\bigotimes_i\ket{N_i=1}\bra{N_i=1}$ in the regime of weak tunneling.

\subsubsection{Tunneling-induced instability of the Mott phase and non-equilibrium phase transition toward a superfluid state}
\label{subsubsec:Mott_instability}
The possibility of stabilizing the Mott Insulating state by mean of a narrowband incoherent pump scheme strongly relies on the presence of nearly flat photonic energy bands. While this approximation is naturally satisfied for weak values $J\ll \Gamma_{\rm{p}}$ of the tunneling constant, more complex behaviours are expected to arise for higher $J\geq\Gamma_{\rm{p}}$: in this case, the kinetic energy shift related to the presence of hole excitations on top of the Mott Insulating state makes some many body-eigenstates to fall outside the spectral range of photonic emission. As a result, the pump is no longer able to refill some of these holes and sustain a commensurate lattice filling. This physical intuition was confirmed by the numerical study in~\cite{Alberto_BH}, which confirmed the presence of a lower density $n<1$ in the high tunneling regime, as well as strong particle number fluctuations and a non-zero entropy. All signatures are a strong indication of a deviation from the ground-state predictions. As we will see in the next section, similar effects related to the presence of flat bands will be involved in the stabilization of Fractional Quantum Hall liquids of light.

In \cite{Alberto_BH}, a mean-field analysis of the microscopic model of Sec.~\ref{subsec:microscopic_model} involving both the emitters and photonic degrees of freedom unveiled that this behavior is associated to a phase transition characterized by a spontaneous breaking of the $U(1)$ symmetry. In analogy with the equilibrium case, this transition can be intuitively understood as the proliferation of hole excitations at some specific kinetic energies favoring delocalization effects and the onset of a long range quantum coherence. Yet, in stark contrast with the usual equilibrium transition from superfluid to Mott insulator state, the transition observed in the photonic model was found to be always triggered by a commensurability effect, and never by the competition between tunneling and the interaction blockade. 


\subsection{Interacting photons in a gauge field: stabilization of Fractional Quantum Hall states of light}\label{subsec:FQH}
In the driven-dissipative context of strongly correlated quantum fluids of light, the first proposal of a non-Markovian reservoir engineering scheme was put forward in~\cite{Kapit} in a circuit-QED framework and regarded the possibility of stabilizing Fractional Quantum Hall states of light in presence of an artificial gauge field.  According to the arguments of Secs.~\ref{subsubsec:flat_Mott} and \ref{subsubsec:Mott_instability}, the presence of a finite many-body energy gap separating the ground state from excited states, along with the presence of flat Landau levels suggests that frequency-dependent incoherent pumping schemes should be able to efficiently stabilize such states.

\subsubsection{Lattice geometry with periodic boundary conditions}
The proposal in~\cite{Kapit} focused on a lattice geometry with periodic boundary conditions in the presence of a synthetic magnetic field for photons. On one hand, the periodic boundary conditions remove the difficulties related to the spurious generation of edge excitations; on the other hand, attention must be paid that a too large dispersion of hole states  induced by the lattice (which would break the flat photonic band assumption) does not hinders the efficiency of the narrowband pumping scheme. To prevent the latter feature, it was suggested in \cite{Kapit} to engineer special lattices architectures first proposed in~\cite{Kapit_Muller} which should mimick the flat Landau levels of the continuum geometry and thus allow to reproduce the FQH physics of the continuous space geometry \cite{Laughlin} at arbitrary filling $\nu=1/k$. 

In \cite{Kapit,Umucalilar_FQH_incoherent}, an important fidelity with the Hamiltonian ground-state reaching values over $95\%$ could be obtained with a narrowband incoherent pump possessing identical properties with respect to the one discussed in Sec.~\ref{subsec:BH_narrwband}. This striking performance can be understood in the following way: for any particle number $N$, all states belonging to the lowest energy submanifold coincide with the lowest Landau level, and the gapped ground-state with $N=N_{\rm{GS}}$ photons corresponds to the Laughlin state. Moreover, in the considered geometry all hole excitations (made of $1/\nu$ quasi-holes) corresponding to states with $N<N_{\rm{GS}}$ are massively degenerate: quasi holes do not have any dispersion nor interact. Finally, states belonging to the higher energy bands, as well as states with $N>N_{\rm{GS}}$, are energetically separated by a gap $\Delta\propto U$ related to the photon-photon interaction blockade. Thus, tuning the Lorentzian emission spectrum on the transition energy required to refill an hole excitation allows for a very efficient refilling toward the Laughlin state, while preventing from populating states with $N>N_{\rm{GS}}$ photons which fall out-of-resonance thanks to the many-body gap $\Delta$.

\begin{figure}[tbph]
\centering
\includegraphics[height=5cm,clip]{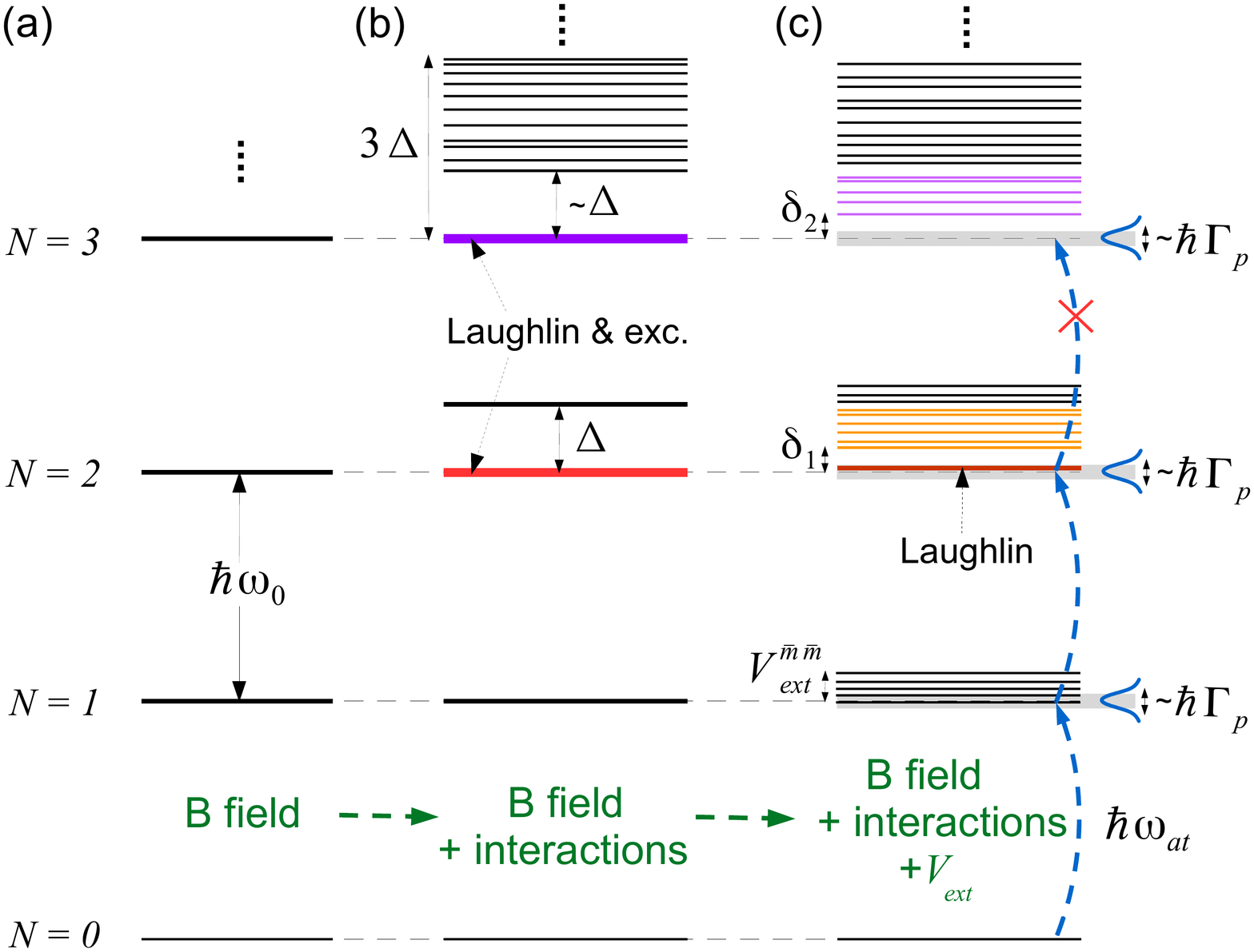} 
\includegraphics[trim=0cm 0cm 70cm 0cm,height=5cm,clip]{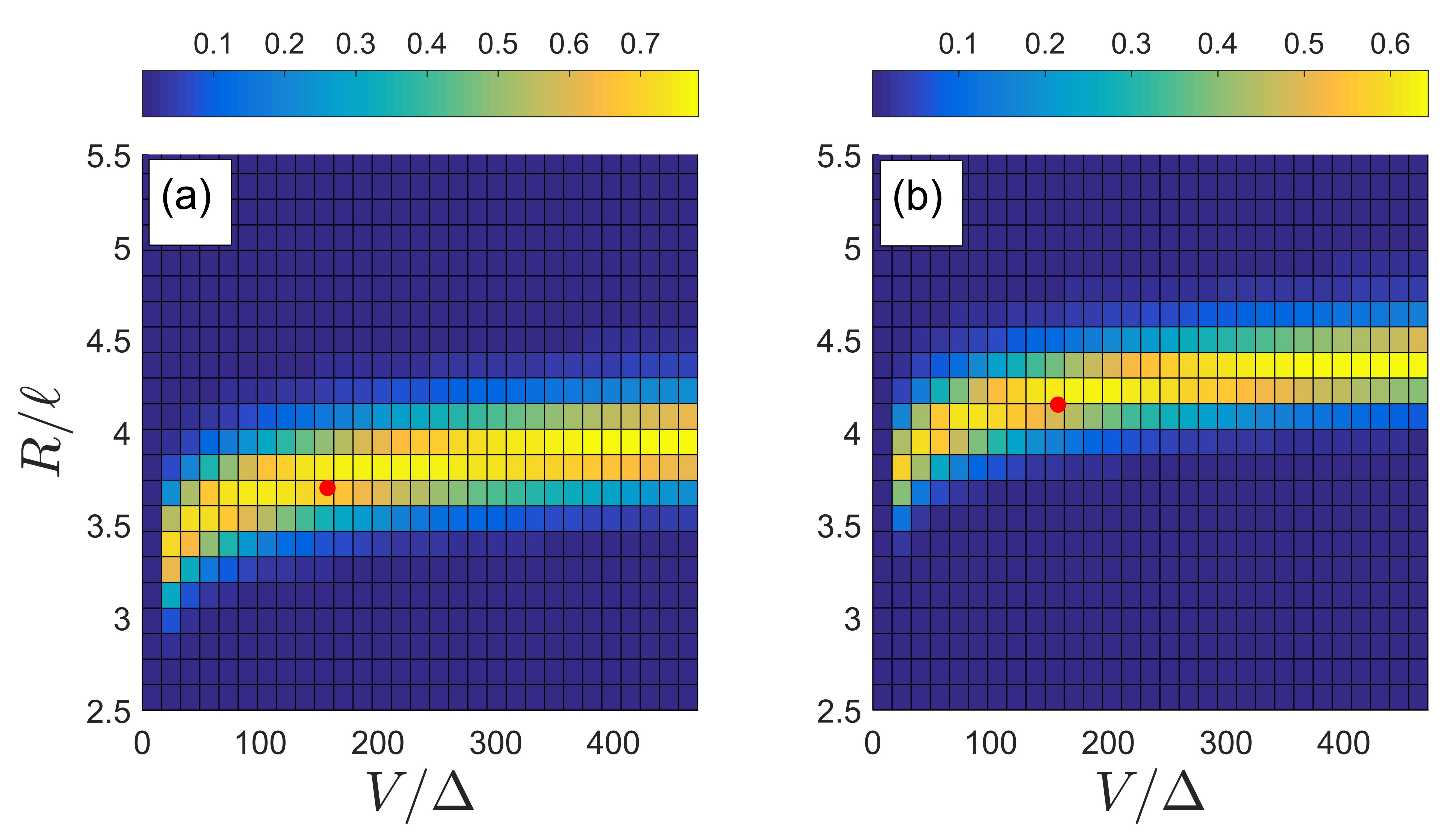} 
\hspace{1cm}
\caption{Left panel: Scheme of the energy levels of an isolated bosonic fractional quantum Hall system.
(a) In the presence of a magnetic field only, noninteracting bosons occupy the lowest Landau level. 
(b) When interactions are added, the Laughlin ground state manifold remains massively degenerate and contains the $\nu = 1/2$ bosonic Laughlin
state, plus its (zero-energy) edge and quasihole excitations. An energy gap of size $\Delta$ separates the Laughlin manifold from bulk excited states.
(c) When an external confining potential is added, the degeneracy of the Laughlin manifold is lifted and all edge and quasihole excitations are raised in energy.
This latter configuration, combined with a narrowband non-Markovian incoherent pumping scheme, allows to selectively generate a fractional quantum Hall state of light. The efficiency of this scheme is displayed in the colorplot shown in the right panel illustrating the Laughlin state fidelity as a function of the parameters of a hard-wall confinement potential. Figure adapted from \cite{Umucalilar_FQH_incoherent}.
\label{fig:fqh}}
\end{figure}
\subsubsection{Continuum space geometry with hard-wall confinment}
While the study of simplified periodic boundary conditions represents a useful step toward a better understanding of the dynamics of driven-dissipative FQH liquids, a realistic experimental implementation would in principle occur in a different geometry and involve the presence of some external confinement. In what follows, we will thus concentrate on the different configuration of a single twisted ring-cavity that has already led to promising experimental observations the strongest experimental interest: as reported in~\cite{Schine_Landau}, this clever system provides a neat quantum simulation of the quantum mechanical motion of a single particle subject to a uniform (synthetic) magnetic field and a harmonic potential, for which the single-particle eigenstates have the form of Landau levels. First studies of interacting particle physics in closely related geometry were then reported in~\cite{Jia_dot}, which displayed promising evidence of photon blockade in the steady-state of the cavity. One can be optimistic about a combination of these two physical effects and a forthcoming realization of fractional quantum Hall physics in this setup.

The application of the incoherent pumping scheme with a narrowband emission spectrum to the generation of Fractional quantum Hall states in this system was studied in~\cite{Umucalilar_FQH_incoherent}. The general structure of the many-body energy levels is displayed in the left panel of Fig.\ref{fig:fqh}: while the excitation of the bulk of the system is protected by a finite energy gap, the edges of the system support low-lying excitations corresponding, for low additional angular momenta, to density waves propagating around the cloud. A complete characterization of these states under different confinement potentials is given in~\cite{elia}. The main difficulty in applying the incoherent pumping scheme to this configuration stems from these edge excitations as we are confronted to the two contradictory constraints: on one hand, the external potential must be able to push the edge excitations out of resonance with respect to the bulk, in such a way to ensure the incompressibility condition and prevent the unconstrained growth of the Laughlin droplet; on the other hand, all edge excitations which might be generated during a photon loss event must remain within the bandwidth of the emitters, so to avoid losing pumping efficiency.

A color plot showing the dependence of the Laughlin state fidelity on the confinement potential parameters is shown in the right panel of Fig.\ref{fig:fqh} for a step-like hard-wall potential of radius $R$ (in units of the magnetic length $\ell$) and height $V$ (in units of the bulk many-body gap $\Delta$). As expected, there exists an interval of position for which both the above-mentioned conditions are well satisfied and the fidelity of the Laughlin state preparation achieves a quite high value. The deviation from one is due to a spurious population left in lower photon number states and to the unwanted excitation of edge states. While the former effect can be tamed by using a larger value of the $\Gamma_{\rm em}^0/\Gamma_l$ ratio, it is not straightforward to design a potential that is able to significanly shift the edge-excited states without affecting the target Laughlin state.

An alternative strategy is to use a more sophisticated square-shaped broadband emission spectra \cite{Lebreuilly_square} instead of Lorentzian narrowband ones: a discussion of this technique and of its remarkable efficiency in the context of Mott insulator states will be presented in the next section, while its application to fractional quantum Hall states is presently the subject of on-going research.

\begin{figure}[t]
\centering
\includegraphics[width=0.7\textwidth,clip]{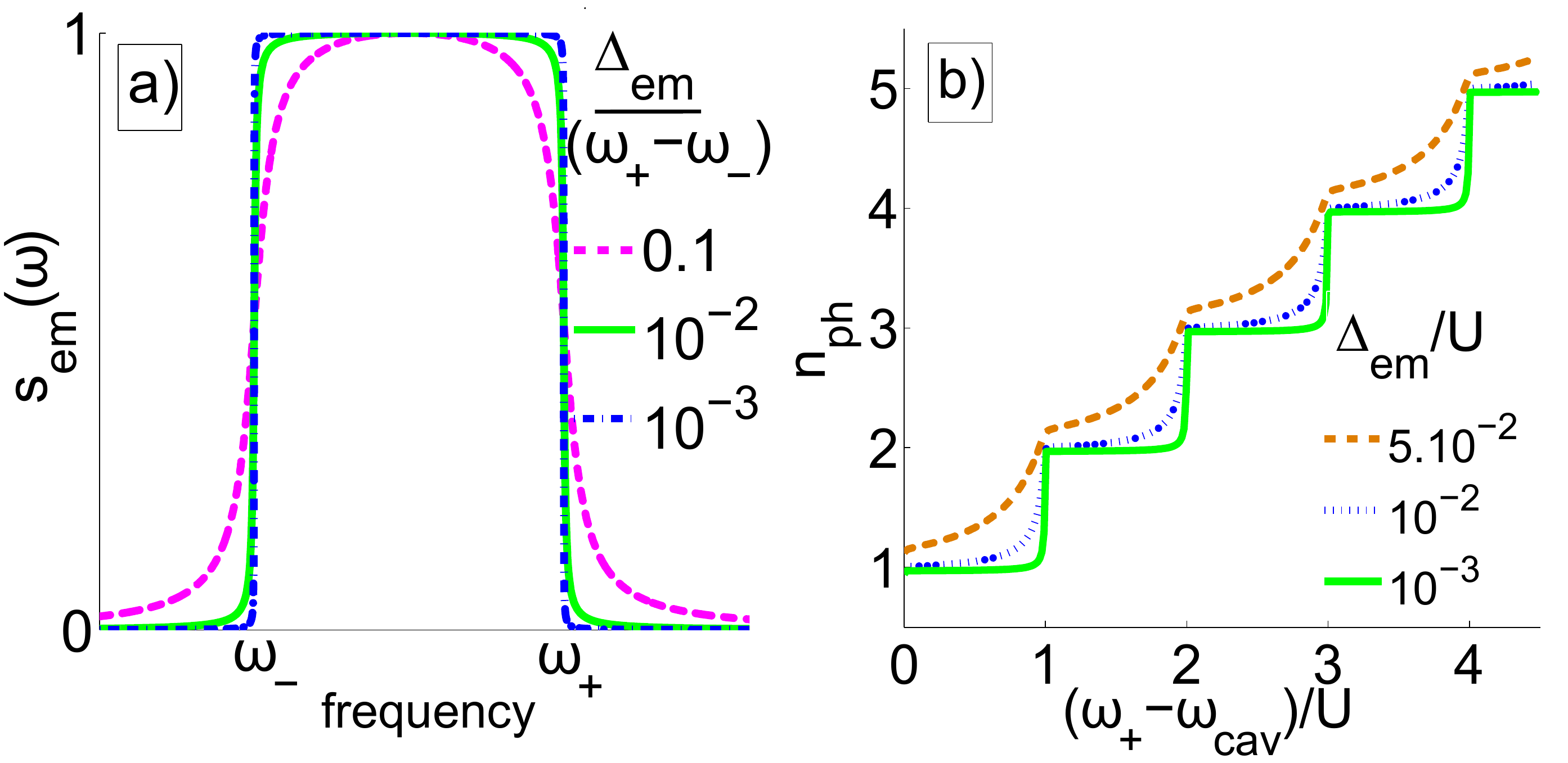}
\caption{\label{fig_square:square-spectrum+plateau} Panel a): Plot of the broadband ``square-shaped'' emission spectrum ${s}_{\rm em}(\omega)=\mathcal{S}_{\rm{em}}(\omega)/\Gamma_{\rm{em}}^0$ defined in Eq.~(\ref{eq:square_emission}) for various values of $\Gamma_{\rm{p}}$ ($\Gamma_{\rm{p}}=\Delta_{\rm{em}}$ within our notation system). Panel b):  Average photon number $n_{\rm{ph}}$  as a function of $\mu=\omega_{+}-\omega_{\rm{cav}}$ (i.e. varying $\omega_{\rm{cav}}$) for a single site system with various values of $\Delta_{\rm{em}}=\Gamma_{\rm{p}}$. Figure adapted from \cite{Lebreuilly_square}.}
\end{figure}

\section{Stabilizing incompressible fluids and zero temperatures states of light via broadband reservoirs}\label{sec:broadband}
In this section we review the physics of strongly interacting photons in contact with broadband reservoirs. Such a configuration was introduced and explored in \cite{Lebreuilly_square}.

In the previous section we have analyzed the potential of a narrowband non-Markovian pumping scheme for the study of strongly correlated states of light. We have confirmed the possibility of stabilizing Mott Insulating and Fractional Quantum Hall photonic states in a regime of flat photonic bands: in the case of the MI states this corresponds to a weak tunneling regime $J/U\ggg1$, while for the FQH case such approach is suited in the presence of flat Landau levels, e.g., in continuous space configurations or some specifically engineered lattices architectures \cite{Kapit_Muller}. On the other hand, important deviations from equilibrium are expected as soon as the photonic bandwidth becomes of the order of the emission linewidth $\Gamma_{\rm{p}}$, which for the BH model corresponds to the regime of higher values of the hopping, and in the continuous space FQH case can naturally arise as a consequence of a trapping potential. In both situations, the deviation from equilibrium is accompanied by a proliferation of many-body excitations on top of the ground state. Even more importantly, we have seen how these deviations profoundly modify the nature of the insulator-to-superfluid transition of a Bose-Hubbard lattice model. In order to use photonic devices to quantum simulate the zero temperature equilibrium physics and its quantum phase transitions, it is therefore necessary to develop more advanced schemes allowing to guide the population towards the Hamiltonian ground-state in a more robust way. As we will show now, a powerful strategy to this purpose is offered by engineered non-Markovian reservoirs with broadband frequency distributions. 

With respect to the case of a narrowband emission, an incoherent pumping scheme with a broadband emission spectrum can be obtained within our proposal of Sec.~\ref{sec:reservoir_engineering} by choosing a collection of emitters with uniformly distributed transition frequencies over the interval $[\omega_-,\omega_+]$, leading to the expression $\mathcal{D}_{\rm{at}}(\omega)=\mathcal{A}\theta(\omega-\omega_{-})\theta(\omega_+-\omega)$, where $\mathcal{A}=N_{\rm{at}}/(\omega_+-\omega_-)$. The resulting emission spectrum, shown in Fig.~\ref{fig_square:square-spectrum+plateau}, panel a), is then of a square shape
\begin{equation}
\label{eq:square_emission}
\mathcal{S}_{\rm{em}}(\omega)=\mathcal{S}_{\rm{em}}^{\rm{Broad}}(\omega)\equiv\Gamma_{\rm{em}}^{0}\mathcal{C}\int_{\omega_-}^{\omega_+}d\tilde{\omega}\frac{(\Gamma_{\rm{p}}/2)^2}{(\omega-\tilde{\omega})^2+(\Gamma_{\rm{p}}/2)^2},
\end{equation}
where $\mathcal{C}$ is a normalizing factor of the integral allowing to the set the maximal value $\mathcal{S}_{\rm{em}}\left((\omega_++\omega_-)/2\right)=\Gamma_{\rm{em}}^0$: $\mathcal{S}_{\rm{em}}(\omega)$ maintains an almost constant value $\Gamma_{\rm{em}}^0$ all over a frequency domain $[\omega_- ,\omega_+]$, and decays smoothly with a power law outside this interval over a frequency scale $\Gamma_{\rm{p}}$. At first sight,  implementing such a pump scheme might appear technologically challenging as it would require in principle a large number of two-level emitters with well-controlled frequencies. Yet, as we have shown in \cite{Lebreuilly_square}, strong hints suggest that only one single emitter with a temporally modulated transition frequency spanning periodically the interval $[\omega_- ,\omega_+]$ would allow to reproduce the effect of such a reservoir, making our proposal within the state-of-the-art of circuit QED.

The main strength of this scheme with respect to the case of the narrow bandpass emission spectrum of last section consists in the possibility of strongly suppressing transitions at high energies, while keeping an efficient pumping on all transitions located at lower frequencies so to efficiently refill the gas without introducing undesired excitations above the many-body band gap. In the case of the BH model, this will allow the Mott state to develop robustness against tunneling, as photons can now be injected across the whole hole band (of width $\sim J$) without emitting undesired excitations above the many-body band gap (of comparable width $\sim U$).

\subsection{Robustness of Mott Insulating states against hopping and losses, and remaining non-equilibrium features}
At the single resonator level, our scheme straightforwardly allows for the stabilization of arbitrary Fock states as shown in Fig.~\ref{fig_square:square-spectrum+plateau}, where we observe a plateau structure in the photon number at steady-state coinciding with that of the equilibrium configuration \cite{Fisher} at zero temperature. In this prospect, the broadband scheme is already more powerful than the previous configuration discussed in Sec.~\ref{sec:narrow_band}, as it allows to select a desired photonic Fock state in a simple manner by playing with the control parameter $\mu=\omega_{+}-\omega_{\rm{cav}}$, e.g, by changing the resonator frequency: in stark contrast with a narrowband scheme, the efficiency of the full stabilization of a given $N_0$-photon Fock state within a given state-of-the-art for the dissipative parameters is roughly independent of the number $N_0$, meaning that Fock state with an high particle number might be as well accessible with a comparable precision.

\begin{figure*}
\includegraphics[width=0.98\textwidth,clip]{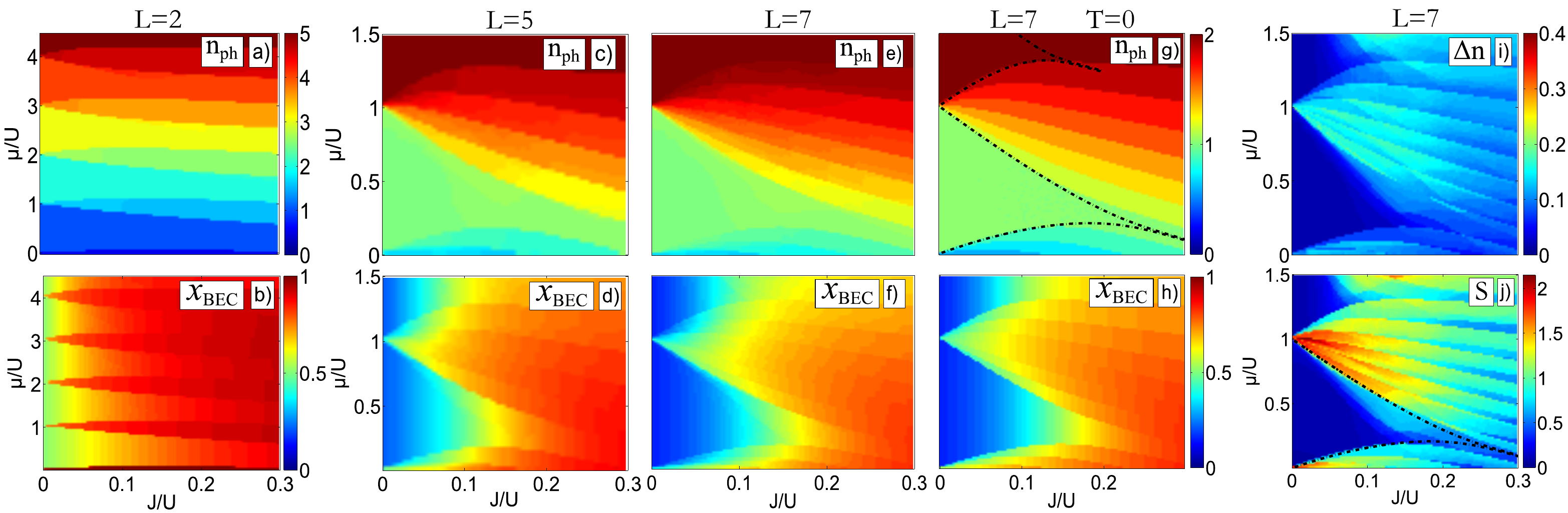}\vspace{-4mm}
\caption{\label{fig_square:phase_diagram}Steady-state properties for a limit choice of parameters allowing to maximize the fidelity with the ground-state predictions. Panels a)-f) show to the steady-state properties, namely average photon number per site $n_{\rm{ph}}$ (upper panels) and the photonic Bose-condensed fraction $x_{\rm{BEC}}$ (lower panels), for various systems sizes $L$ (namely $L=2,5,7$ from left to right). Panel g) and h) respectively shows the average $n_{\rm{ph}}$  and $x_{\rm{BEC}}$ in a $T=0$ equilibrium system for $L=7$. Panel i) and j) respectively show the steady-state particle number relative fluctuations $\Delta n$  and the entropy $S=-\langle \text{ln}(\rho_{\infty})\rangle$], for $L=7$. In panel g) and j)  dash dotted black lines indicate the MPS $T=0$ prediction for the first Mott lobes.  The choice of parameters is done in such a way to maximize the ground-state occupancy. Figure adapted from \cite{Lebreuilly_square}
}
\end{figure*}

Pushing the analogy with equilibrium further, many arguments strongly support that even in the extended lattice configuration the steady-state of the driven-dissipative model Eq.~(\ref{eq:photon_only}) should present an important overlap with a zero temperature equilibrium state of chemical potential $\mu=\omega_{+}-\omega_{\rm{cav}}$, i.e., with the ground-state $\ket{GS}$ of the effective Hamiltonian
\begin{equation}
\label{eq:effective_hamiltonian}
H_{\rm eff}=\sum_{i=1}^{L}\left[-\mu a_{i}^{\dagger}a_{i}+\frac{U}{2}a_{i}^{\dagger}a_{i}^{\dagger}a_{i}a_{i}\right]{-}\sum_{\avg{i,j}}Ja_{i}^{\dagger}a_{j}.
\end{equation}
This is confirmed in Fig.~\ref{fig_square:phase_diagram}, where we plot the steady-state properties [Panels a)-f)] for a periodic 1D chain and for several system sizes $L$, and compare them with the zero temperature predictions [Panels g),h)]: the steady-state presents a series of lobes featuring successive integer values of the photonic density $n_{\rm{ph}}=\langle N\rangle/L$ (see upper panels) , accompanied by weak values of the Bose-condensed fraction $x_{\text{BEC}}=\langle n_{k=0}\rangle/\langle N\rangle$, indicating a spatial localization effects of photons within the system. The remaining non-vanishing value $x_{\text{BEC}}\sim 1/L$ for $J/U\to 0$ is a simple consequence of the finite system size, also visible in the equilibrium case (Panel h), meaning that the localization effect has reached the ultimate limit in which each photon is pinned on a single site, and that the various resonators are completely decoupled: $\rho_{\infty}=\bigotimes_i\ket{N_i=N_0}\bra{N_i=N_0}$. Moreover, as shown in panel i), a large part of the Mott region features strongly suppressed relative fluctuations $\Delta n\equiv\sqrt{\langle N^2\rangle-\langle N\rangle^2}/\langle N \rangle\simeq 10^{-2}$ in the total photon number $N=\sum_i a_i^\dagger a_i$: this is a  non-equilibrium form of incompressibility. As anticipated, a very good agreement with the ground-state properties of $H_{\rm{eff}}$ can be observed by comparing panels e) and f) to panels g) and h) which respectively features the steady-state and zero temperature predictions for the same size $L=7$: in particular the boundary of the Mott region of the dissipative model (shown, e.g., in panel e)) follows closely the prediction for the phase boundary toward the superfluid region (shown in black dashed line in panel g)) obtained in the thermodynamic limit of the equilibrium model by mean of matrix product state simulations with $L=200$. The small differences between the two domains, which can be also seen in the equilibrium system for $L=7$ [panel g)], are likely to be finite-size effects.

\begin{figure}[t]
\begin{center}
\includegraphics[width=0.7\textwidth,clip]{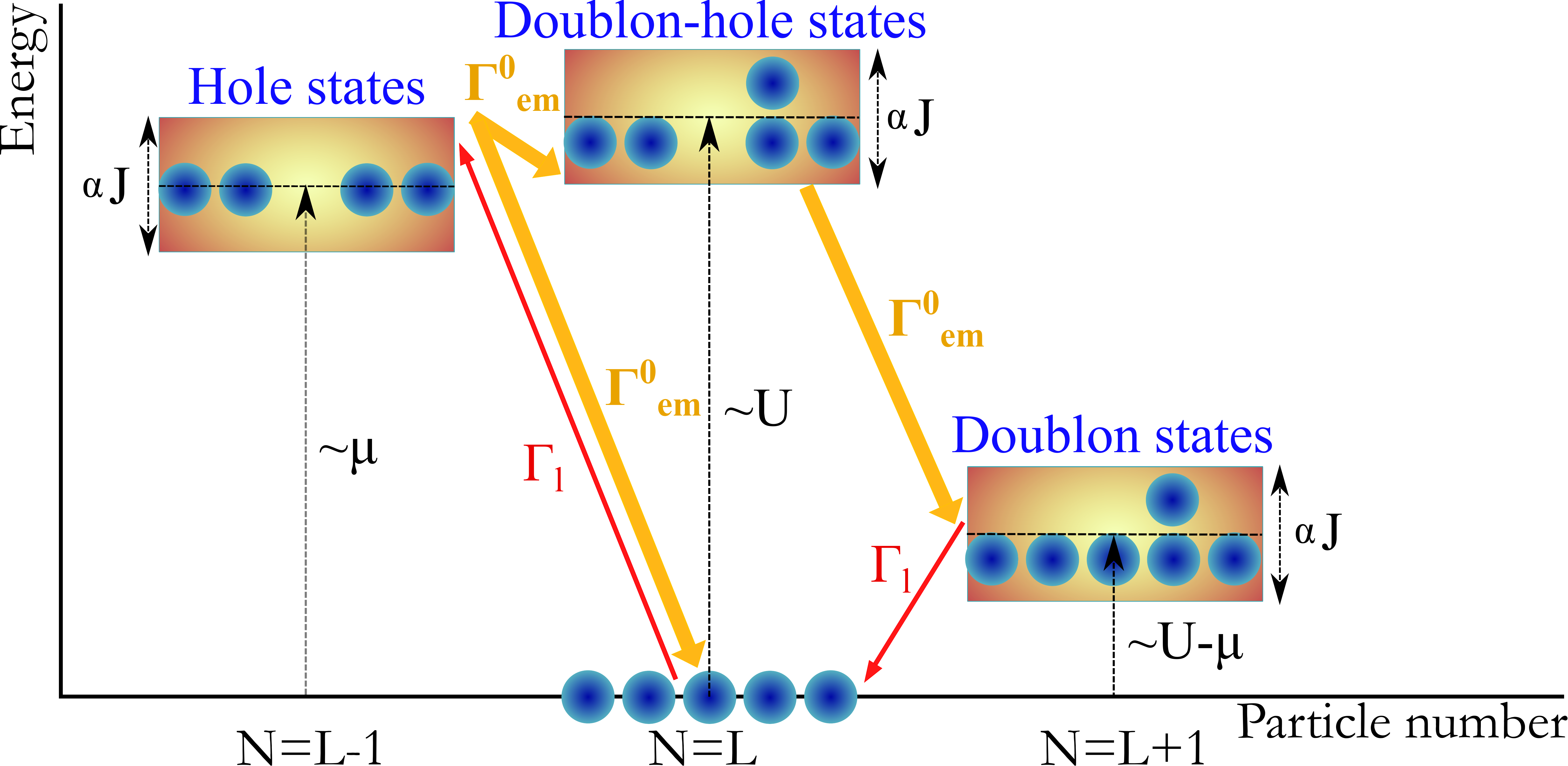}
\end{center}
\caption{\label{fig:doublon} Sketch of the non-equilibrium scenario in the thermodynamic limit: above a critical hopping $J_{c}$, the hole and doublon-hole energy bands start overlapping and are thus coupled by emission process (which are only enhanced for transition diminuishing the total energy computed with $H_{\rm{eff}}$). This allows for a probability leakage toward doublon states after a photon loss event. Going back towards the ground-state requires an additional slow photon loss event. Adapted from \cite{Lebreuilly_PhD}}
\end{figure}

In virtue of this strong resemblance with the zero temperature prediction, a tempting conjecture would be that the steady-state indeed maps onto the ground-state of $H_{\rm{eff}}$. However, some striking deviations can be observed in the panel j) of Fig.~\ref{fig_square:phase_diagram} which shows the steady-state entropy $\mathcal{S}=-\text{Tr}\left[\rho_{\infty}\rm{ln}(\rho_{\infty})\right]$ of the driven-dissipative model: while a large part of the Mott regions features values of $\mathcal{S}$ close to zero as this is expected from a zero temperature state, outside these domains a sharp increase of $\mathcal{S}$ is observed, which indicates the presence of a statistical mixture and is a strong criterion of departure from $T=0$. We precise that the $\mathcal{S}\simeq 0$ regions, which feature an almost pure quantum state at steady-state, also exhibited a very high fidelity $\mathcal{F}\equiv \bra{GS}\rho_{\infty}\ket{GS}\geq 0.99$ with the ground-state of $H_{\rm{eff}}$, meaning that in this range of parameters our scheme indeed quantum simulates the zero temperature physics. The departure from equilibrium marked by the emergence of entropy can be understood as a non-equilibrium process occuring over some critical value $J_{c}$ for the hopping terms and inducing some probability leakage from the ground-state. Such process, schematized in Fig.~\ref{fig:doublon} for $J\geq J_{c}$, involves higher energy bands as catalysers and leads to the generation of long-lived doublon excitations. For $J<J_{c}$ the high energy bands do not overlap and the ground-state is dynamically well protected.

\begin{figure}[t]
\centering
\includegraphics[width=0.5\textwidth]{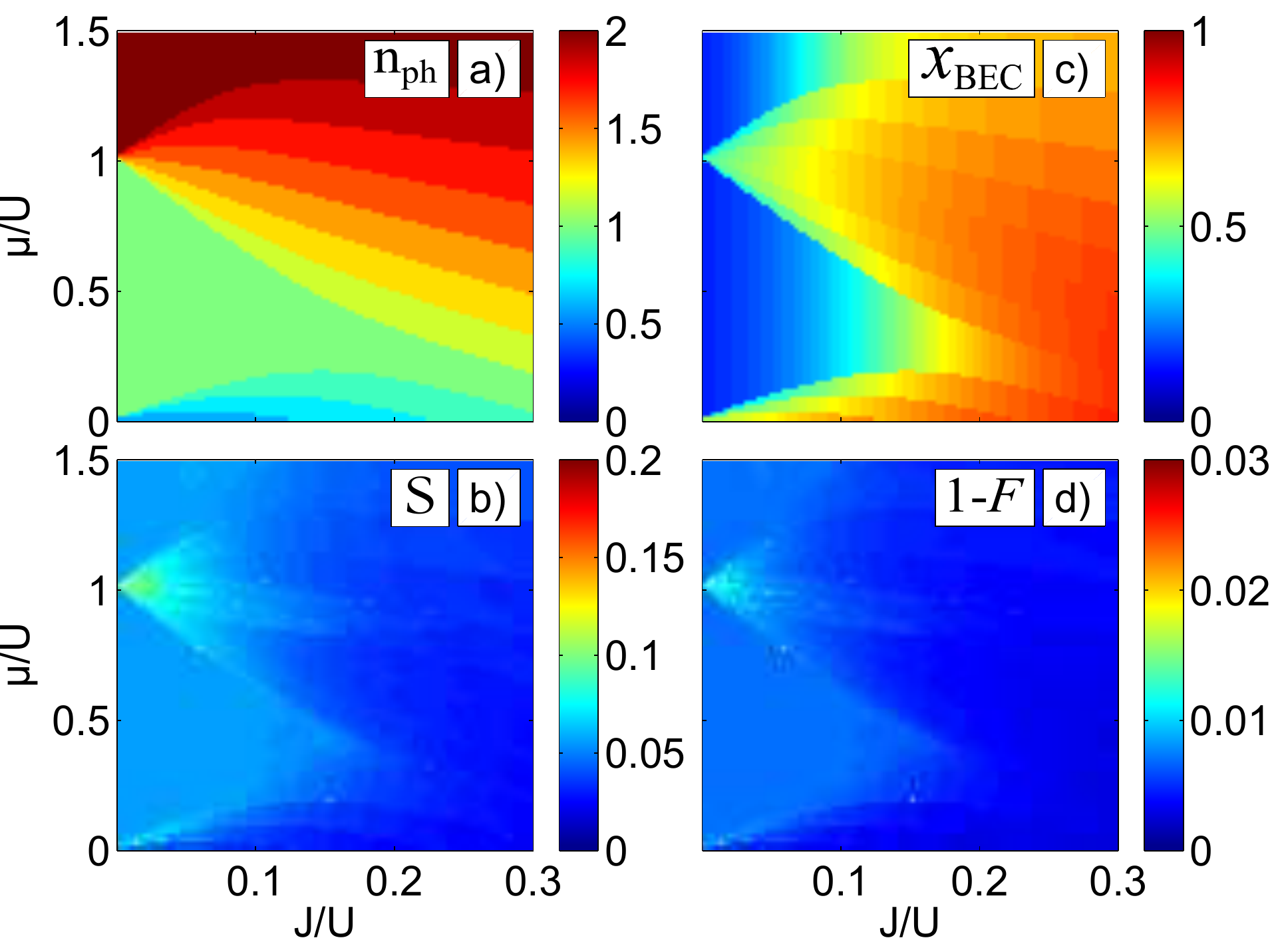}
\caption{\label{fig:frequency_dependent_losses}Steady-state properties in the presence of frequency-dependent losses with a square-spectrum in addition to a frequency-dependent emission. Simulations were done for a $L=7$ sites periodic chain. Panel a) (resp. panel c): Average steady-state photon number per site $n_{\rm{ph}}$ (resp. condensed fraction $x_{\rm{BEC}}$). Panel b): steady-state entropy. Panel d): $1-\mathcal{F}$, where $\mathcal{F}=\bra{GS}\rho_\infty\ket{GS}$ is the fidelity of the steady-state density matrix $\rho_{\infty}$ with the ground-state $\ket{GS}$ of the Hamiltonian $H_{\rm{eff}}$, i.e, a $T=0$ state of chemical potential $\mu$. The choice of parameters is done in such a way to maximize the ground-state occupancy. Figure adapted from \cite{Lebreuilly_square}.}
\end{figure}

\subsection{Implementation of additional frequency-dependent losses: full ground-state quantum simulation}
\label{subsec:frequency-dependent_losses}
With respect to the narrowband case, the use of a broadband non-Markovian reservoir for the photonic emission already allows to reach a much higher performance and fidelity with the ground-state prediction. Yet, the previously mentioned mechanism leading to the generation of a residual entropy might be seen as a remaining hindrance in the prospect of quantum simulating zero temperature physics in photonic lattices. We now review our results regarding a further extension of the optical scheme, based on the use of frequency-dependent losses that is able to remove this deviation from equilibrium. More precisely, we consider the following dynamics for the photonic density matrix
\begin{equation}
\partial_{t}\rho(t) =  -i\left[H_{\rm ph},\rho(t)\right]+\mathcal{L}_{\rm{l}} \big[\rho(t)\big]+ \mathcal{L}_{\rm{em}}\big[\rho(t)\big]+\mathcal{L}^{(\text{add})}_{\rm{L}} \big[\rho(t)\big],
\label{eq_square:photon_only2}
\end{equation}
where the Hamiltonian and dissipative contributions $H_{\rm ph}$, $\mathcal{L}_{\rm{l}} \big[\rho(t)\big]$ and $\mathcal{L}_{\rm{em}}\big[\rho(t)\big]$ are left unchanged with respect to the beginning of Sec.~\ref{sec:broadband}. Similarly to emission, the additional frequency-dependent loss term
\begin{equation}
\label{eq_square:loss-non-markov}
\mathcal{L}^{(\text{add})}_{\rm{L}} \big[\rho(t)\big] = \frac{\Gamma_{\rm{L}}^{0}}{2}\sum_{i=1}^{L}\left[\bar{a}_{i}\rho a_{i}^{\dagger}+a_{i}\rho\bar{a}_{i}^{\dagger}-a_{i}^{\dagger}\bar{a}_{i}\rho-\rho\bar{a}_{i}^{\dagger}a_{i}\right]\!.
\end{equation}
involves modified lowering ($\bar a_i$) and raising ($\bar a^\dagger_i\equiv [\bar a_i]^\dagger $) operators
\begin{equation}
\label{eq_square:special-operators2}
\frac{\Gamma^{0}_{\rm{L}}}{2}\bar{a}_{i} = \!\int_{0}^{\infty}d\tau\, \Gamma_{\rm{L}}(\tau)a_{i}(-\tau),
\end{equation}
where 
\begin{equation}
\label{eq_square:memory-kernel2}
\Gamma_{\rm{L}}(\tau)=\theta(\tau)\int \frac{d\omega}{2\pi}\mathcal{S}_{\rm{L}}(\omega)e^{-i\omega\tau},
\end{equation}
and
\begin{equation}
\mathcal{S}_{\rm{L}}(\omega)=\Gamma_{\rm{L}}^{0}\mathcal{C}'\int_{\omega_+}^{\omega_{\rm{L}}}d\tilde{\omega}\frac{(\Gamma_{\rm{p}}/2)^2}{(\omega-\tilde{\omega})^2+(\Gamma_{\rm{p}}/2)^2},
\label{eq_square:square_spectrum2}
\end{equation}
is the frequency-dependent loss rate, which  we also choose to be of a square shape as the emission term.
$\mathcal{C}'$ is a normalizing factor of the integral allowing to the set the maximal value $\mathcal{S}_{\rm{L}}\left((\omega_{\rm{L}}+\omega_+)/2\right)=\Gamma_{\rm{L}}^0$. In analogy with the emission term, frequency-dependent losses with a broadband profile can be obtained by coupling our system to absorbers with transition frequencies uniformly distributed over $[\omega_{+}, \omega_{\rm{L}}]$ and a strong dissipative decay $\Gamma_{\downarrow}=\Delta_{\rm{L}}$ toward the ground-state.

In consequence of our non-Markovian reservoir engineering procedure, both non-Markovian emission and loss processes are now configured in such a way to strongly enhance transitions between many-body eigenstates which reduce the total energy computed with the Hamiltonian $H_{\rm{eff}}$ of Eq.~(\ref{eq:effective_hamiltonian}). Thus, the only well-isolated and dynamically-protected eigenstate is the ground-state $\ket{GS}$ of $H_{\rm{eff}}$ (with $N_{\rm{tot}}$ photons), since it does not have states with $N_{\rm{tot}}-1$ and $N_{\rm{tot}}+1$ photons with lower energy: while a photon might still leak out of the system from time to time because of a natural loss process, an energy decay dissipative process involving both frequency-dependent engineered emission and losses should then bring back probability into the Hamiltonian ground-state at a much higher rate. In particular the entropy generation processes leading to generation of long-lived doublon excitations which we mentioned earlier can not occur with this new scheme. The efficiency of this new method is confirmed in Fig.~\ref{fig:frequency_dependent_losses}, as the average values of simple observables $n_{\rm{ph}}$ [panel a)] and $x_{\rm{BEC}}$ [panel b)] are now completely undistinguishable from the $T=0$ prediction (see panels g) and h) of Fig.~\ref{fig_square:phase_diagram}). More importantly, very weak values $\mathcal{S}\leq 0.1$ of the steady state entropy  (panel c) of Fig.\ref{fig:frequency_dependent_losses}) and a full fidelity $\mathcal{F}=\bra{GS}\rho_{\infty}\ket{GS}\geq 0.99$ [panel d)] with the many-body ground-state are now observed everywhere in the parameter space. In contrast with the original scheme, there appears to be no real physical limitations to how close the steady-state can be to the ground-state $\ket{GS}$, as we have checked that these remaining corrections were a mere consequence of the finite choice of the dissipative parameters and could be arbitrarily reduced.

The fact that this improved scheme succeeds to stabilize the ground-state of the Bose-Hubbard model both in the Mott insulating and superfluid regimes independently of the details of the underlying many-body physics (which is significantly different in the $J\ll U$ or $J\gg U$ cases) and across the phase transition separating the Mott insulator and the superfluid regimes is a strong indication of its robustness and flexibility. We are therefore confident that this scheme can be efficiently applied to the quantum simulation of the zero temperature physics of a much wider range of Hamiltonians. 
 
\section{Conclusions and future directions}
\label{sec:conclusions}

In this article we have briefly reviewed new non-Markovian reservoir engineering schemes designed  to generate and stabilize interesting quantum many-body states of a fluid of strongly interacting photons. The idea is to exploit the frequency-dependence of the emission implemented, e.g., via population-inverted two-level emitters, so to selectively inject the photons into the desired many-body state and simultaneously suppress the population of excited states. The strong promise of this scheme is illustrated on the most celebrated cases of fractional quantum Hall  and Mott insulator states of light and of the associated phase transitions to superfluid states. Depending on the properties of the chosen artificial reservoirs, the steady-state can on demand present exotic non-equilibrium signatures or alternatively reproduce the Hamiltonian equilibrium ground-state properties with an high level of fidelity.

Remaining in that quantum simulation perspective, we have seen how the use of reservoirs with broadband spectra allows for the stabilization of the many-body ground state across the Mott insulator to superfluid phase transition.  A direct next step will be to extend this study in the presence of synthetic gauge fields, so to overcome the limitations of existing schemes and increase the fidelity of fractional quantum Hall states of light. In this last context, a challenging question will be to investigate the behaviour of our stabilization schemes in the presence of multiple ground states and assess the possibility of a topological protection of the quantum bit encoded in the degenerate manifold. In connection with Schr\"{o}dinger cat states which are attracting a growing attention in the quantum fluids of light community \cite{Bartolo_cat,Savona_cat} and hold strong promises of quantum computation applications \cite{Mirrahimi_cat,Leghtas_cat}, it would be exciting to look for more exotic ground-states, e.g. spontaneously breaking a $U(1)$ symmetry and generating a superfluid order that however preserves a $Z_2$ symmetry.

Future challenges will also regard the problem of further simplifying the proposed experimental scheme and finding ways to reproduce tailored dissipative spectra without any drastic increase of the number of emitter families. Along similar lines, there is an interesting fondamental problem of transport of photons from the emitters throughout the rest of the system: while many works \cite{Ma_Simon,Lebreuilly_square} already support the idea that for a small chain it is indeed sufficient to couple only a few sites to the engineered reservoirs, it is in fact not clear how the steady-state properties depend in the generic case on the spatial distribution of emitters and absorbers, and whether new phases of the photon fluid can appear if the density of emitting sites is too small to stabilize a spatially homogeneous state.

From the angle of non-equilibrium quantum dynamics, the dynamical behaviour of the system in the presence of relatively strong pumping and dissipation is also raising intriguing questions. For instance, it will be of great interest to quantify how much the difference from equilibrium depend on the bath coupling strength and how the driven-dissipative nature will affect the collective modes of the system. While the appearance of a finite effective temperature is a quite ubiquitous consequence of dissipation, what happens is not clear when the non-hermitian pumping and dissipation becomes comparable to the Hamiltonian dynamics, so that strong coupling effects between the photon fluid and the reservoir may become possible and lead to new entangled degrees of freedom.

Finally, reformulating our work in a quantum information langage, the broadband configuration can also be seen as an explicit realization of quantum annealing, based however on a differing approach with respect to the traditional adiabatic quantum computation (AQC) methods~\cite{Quantum_annealing,Quantum_annealing_2}, as it is here the interplay between the Hamiltonian dynamics and the relaxation mechanisms related to pumping and losses who progressively pushes the many-body system towards a (possibly strongly entangled) state minimizing energy. Yet, many arguments~\cite{Kechedzhi_annealing,Nishimura_annealing,Quantum_annealing_relaxation} strongly support the fact that an hybrid scheme mixing both aspects of the AQC protocol and the dissipative relaxation scheme would allow to release some severe adiabaticity constraints and fasten the computation. It would be thus interesting to assess the efficiency of non-Markovian reservoir engineering techniques in the context of optimization problem solving.

While the results reported in this article were obtained using relatively elementary techniques, going deeper into the non-equilibrium many-body dynamics will likely require the use of more sophisticated tools to cope with the complex interplay of strong interactions, pumping and dissipation in a non-Markovian context. Numerical mean-field techniques and Keldysh diagrams are of course powerful choices, but the variety of elementary excitations and the richness of the expected phase diagram appear to be well worth the effort of looking for completely new theoretical avenues.

\section{Acknowledgements}
\label{sec:Acknowledgements}

The work reviewed in this article was carried out in continuous collaboration with Alberto Biella, Florent Storme, Davide Rossini, Rifat Onur Umucal\i lar, Michiel Wouters, Rosario Fazio and Cristiano Ciuti. We are also grateful to Mohammad Hafezi, Jonathan Simon, Elia Macaluso, Alessio Chiocchetta, Tomoki Ozawa, and Hannah M. Price for stimulating exchanges. This work was supported by Provincia Autonoma di Trento, partly through the SiQuro project (``On Silicon Chip Quantum Optics for Quantum Computing and Secure Communications''), from ERC through the QGBE grant and from the EU-FET Proactive grant AQuS, Project No.640800 and the EU-FET-Open grant MIR-BOSE Project No.737017.


\end{document}